\documentclass[preprint,notoc]{JHEP3} % 10pt is ignored!
\usepackage{epsfig,multicol}
\usepackage{packfig10}
\usepackage{amssymb}
\usepackage{graphics}

\usepackage{amsmath,latexsym}
\newcommand{\ii}{\'\i}

\def\beq{\begin{eqnarray}}    %%%  begequation/eqnarray
\def\eeq{\end{eqnarray}}      %%%  endequation/eqnarray

\newcommand{\OM}{\Omega_M}
\newcommand{\OL}{\Omega_{\Lambda}}

\newcommand{\OK}{\Omega_K}
\newcommand{\OZ}{\Omega_0}
\newcommand{\OT}{\Omega_T}
\newcommand{\rc}{\rho_c}
\newcommand{\rM}{\rho_M}
\newcommand{\rR}{\rho_R}
\newcommand{\CC}{\Lambda}
\newcommand{\bCC}{\beta_{\Lambda}}

\title{Testing the running of the cosmological constant with
 Type I{\rm a} Supernovae at high z}
\author{Cristina Espa\~{n}a-Bonet, Pilar
Ruiz-Lapuente\\
Department of Astronomy, and CER for Astrophysics, Particle
Physics and Cosmology, University of Barcelona, Diagonal 647,
E-08028, Barcelona, Spain \\ 
E-mail: cespana@am.ub.es and pilar@am.ub.es}
\author{Ilya L. Shapiro,\\
Departamento de Fisica, ICE, Universidade Federal de Juiz de
Fora, MG, Brazil\\ E-mail: shapiro@fisica.ufjf.br}
\author{Joan Sol\`{a}\\
Departament d'Estructura i Constituents de la Mat\`eria, and CER
for Astrophysics, Particle Physics and Cosmology, Universitat de
Barcelona, Diagonal 647, E-08028, Barcelona, Spain, and Institut
de F\'{\i}sica d'Altes Energies, E-08193, Bellaterra, Barcelona,
Spain. E-mail: sola@ifae.es}

\preprint{ UB-ECM-PF-03/06}

\abstract{ Within the Quantum Field Theory context the idea of a
``cosmological constant'' (CC) evolving with time looks quite
natural as it just reflects the change of the vacuum energy with
the typical energy of the universe. In the particular frame of
Ref. \cite{Letter}, a ``running CC'' at low energies may arise
from generic quantum effects near the Planck scale,  $M_P$,
provided there is a smooth decoupling of all massive particles
below $M_P$. In this work we further develop the cosmological
consequences of a ``running CC'' by addressing the accelerated
evolution of the universe within that model. The rate of change of
the CC stays slow, without fine-tuning, and is comparable to
$H^2\,M_P^2$. It can be described by a single parameter, $\nu$,
that can be determined from already planned experiments using SNe
Ia at high z. The range of allowed values for $\nu$ follows mainly
from nucleosynthesis restrictions. Present samples of SNe Ia can
not yet distinguish between a ``constant'' CC or a ``running''
one. The numerical simulations presented in this work show that
SNAP can probe the predicted variation of the CC either ruling out
this idea or confirming the evolution hereafter expected.}


\begin{document}

%%%%%%%%%%%%%%%%%%%%%%%%%%%%%%%%%%%%%%%%%%%%%%%%%%%%%%%%%%%
\section{Introduction}

The Standard Cosmological Model fits our universe, in the large,
into an homogeneous and isotropic
Friedmann-Lema\^{\i}tre-Robertson-Walker (FLRW) cosmological
type\, \cite{Peebles}. Its 4-curvature is determined from the
various contributions to its total energy density, namely in the
form of matter, radiation and cosmological constant.

Evidence for a dominant content of energy in the form of
cosmological constant was found by tracing the rate of expansion
of the universe along z with high--z Type Ia supernovae \cite{p99,
Riess98}. This measurement combined with the measurements of the
total energy density $\OT^0$ from the CMB anisotropies
\cite{CMBR,WMAP}, indicates that $\OL^0\sim 70\%$ of the critical
energy density of the universe is cosmological constant (CC) or a
dark energy candidate with  a similar dynamical impact on the
evolution of the expansion of the universe. The matter content,
on the other hand is dominated by the dark matter, whose
existence is detected by dynamical means \cite{Peebles}, and
amounts to $\OM^0\sim 30\%$ of the critical density.

The CC value found from Type Ia supernovae at high z
\cite{p99,Riess98} is:
\begin{equation}\label{CCvalue}
\CC_0=\OL^0\,\rc^0\simeq 6\,h_0^2\times 10^{-47}\,GeV^{4}\,.
\end{equation}
{Here $\rc^0 \simeq \left( 3.0\,\sqrt{h_{0}}\times
10^{-12}\,GeV\right) ^{4}$ is the present value of the critical
density, and $h_{0}\sim 0.7\pm 0.1$\ sets the typical range for
today's value of Hubble's constant $H_{0}\equiv 100\;(Km/sec\,
Mpc)\;h_{0}$}. In the context of the Standard Model (SM) of
electroweak interactions, this measured CC should be the sum of
the original vacuum CC in Einstein's equations,
 $\CC _{vac}$, and the induced
contribution from the vacuum expectation value of the Higgs
effective potential, $\CC_{ind}=\langle V_{\rm eff} \rangle$:
\begin{equation}
\CC=\CC _{vac}+\CC _{ind}\,. \label{lambdaphys}
\end{equation}
It is only this combined parameter that makes physical sense,
whereas both $\CC _{vac}$ and $\CC _{ind}$ remain individually
unobservable\footnote{{In general the induced term may also get
contributions from strong interactions, the so-called quark and
gluon vacuum condensates. These are also huge as compared to
(\ref{CCvalue}), but are much smaller than the electroweak
contribution $V_{\rm eff}$.}}. From the current LEP 200 numerical
bound on the Higgs boson mass, $\,M_{H}> 114.1\,GeV$
\cite{LEPHWG}, one finds $\,\,\,\left| \CC
_{ind}\right|=(1/8)M_H^2\,v^2> 1.0\times 10^{8}\,GeV^{4}$, where
$v\simeq 246\,GeV$ is the vacuum expectation value of the Higgs
field. Clearly, $\left| \CC _{ind}\right|$ is $55$ orders of
magnitude larger than the observed CC value (\ref{CCvalue}). Such
discrepancy, the so-called
 ``old'' cosmological constant problem
 \cite{weinRMP,CCRev}, manifests itself in the necessity of
enforcing an unnaturally exact fine tuning of the original
cosmological term $\CC _{vac}$ in the vacuum action that has to
cancel the induced counterpart $\,\CC _{ind}$ within a precision
(in the SM) of one part in $\,10^{55}$.

The measured CC remains very small as compared to the huge CC
value predicted in the SM of Particle Physics. {Actually, if the
physical value of the CC would conform with that one predicted in
the SM, the curvature of our universe would be so high that the
Special Theory of Relativity could not be a solution to Einstein
equations to any reasonable degree of approximation. Therefore,
the SM prediction of the CC violently contradicts our experience,
whereas the small measured value (\ref{CCvalue}) is perfectly
compatible with it.

The Cosmological Constant Problem (CCP) is a fundamental problem.
It is most likely related to the delicate interplay between
Gravity and Particle Physics, and it has become one of the main
poles of attention \cite{weinRMP,CCRev}. All attempts to deduce
the small value of the cosmological constant from a sound
theoretical idea ended up with the necessity of introducing severe
fine-tuning. This concerns also, unfortunately, the use of
supersymmetry and string theory (see e.g.
\cite{wittenDM,Carroll2}). In this respect we recall that, for a
realistic implementation of the existing versions of M-Theory,
one would like to have a negative (or at least vanishing)
cosmological constant in the remote future, such that it does not
prevent the construction of the asymptotic S-matrix states in
accelerated universes\,\cite{MTheory}. {Since the presently
observed value of the CC is positive, there is the hope that a
variable cosmological term may solve this problem.}

{There is a permanently growing flux of proposals concerning the
CCP.} On the first place there is the longstanding idea of
identifying the dark energy component with a dynamical scalar
field \cite{Dolgov,PSW}. More recently this approach took the
popular form of a ``quintessence'' field slow--rolling down its
potential\,\cite{Caldwell}. This proposal has, on its own, given
rise to a wide variety of models \cite{PeebRat,Moreq}. Extended
models of this kind (``k--essence'') are also based on scalar
fields but with a non-canonical kinetic energy \cite{kessence}.
{The main advantage of the quintessence models is that they could
explain the possibility of a variable vacuum energy.} This may
become important in case such variation will be someday detected
in the observations. Recently other approaches have appeared in
which the dark energy is mimicked by new gravitational physics
\cite{Carroll}. From the point of view of the CCP, all these
approaches lead {to the introduction of either a very small
parameter} or a very high degree of fine-tuning. In another, very
different, vein the possibility to accept the observed value of
the CC within the context of a many world pool is offered by the
anthropic proposal \cite{antrop}. Let us finally mention the
intriguing proposal of non-point-like gravitons at sub-millimeter
distances suggested in \cite{Sundrum2}, or the possibility of
having multiply degenerate vacua\,\cite{Yokoyama}.

When assessing the possibility to have variable dark energy,
other no less respectable possibilities should be taken into
account. In a series of recent papers \cite{JHEPCC1,cosm}, the
idea has been put forward that already in standard Quantum Field
Theory (QFT) one would not expect the CC to be constant, because
the Renormalization Group (RG) effects may shift away the prescribed
value, in particular if the latter is assumed to be zero.  Thus,
in the RG approach one takes a point of view very different from
e.g. the quintessence proposal, as we deal all the time with a
``true'' cosmological term. It is however a variable one, and
therefore a time-evolving, or redshift dependent: $\CC=\CC (z)$.
Although we do not have a QFT of gravity where the running of the
gravitational and cosmological constants could ultimately be
substantiated, a semiclassical description within the well
established formalism of QFT in curved space-time (see e.g.
\cite{birdav,book}) should be a good starting point.  From the RG
point of view, the CC becomes a scaling parameter whose value
should be sensitive to the entire energy history of the universe
-- in a manner not essentially different to, say, the
electromagnetic coupling constant. One of the main distinctions
between our approach and all kinds of quintessence models is that
these models imply the introduction of a phenomenological
equation of state $p_{\chi}=w\,\rho_{\chi}$ for the scalar field
$\chi$ mimicking the CC, where $w$ is a negative index (smaller
than $-1/3$). Whether constant or variable,  a ``true''
cosmological parameter has, instead, no other equation of state
associated to it apart from the exact $w=-1$ one.

Attempts to apply the RG for solving the CC problem have been
made earlier \cite{Polyakov,lam}. The {canonical form} of
renormalization group equation (RGE) for the $\CC$ term at high
energy is well known -- see e.g. \cite{book,Nelson}. {However, at
low energy decoupling effects of the massive particles may change
significantly the structure of this RGE, with important
phenomenological consequences.} This idea has been retaken
recently by several authors from various interesting points of
view \cite{JHEPCC1,cosm,Babic,Reuter03a}. However, it is not easy
to achieve a RG model where the CC runs smoothly without fine
tuning at the present epoch. {In Ref.\,\cite{Letter,IRGA03} a
successful attempt in this direction has been made, which is
based on possible quantum effects near the Planck scale.} At the
same time, the approximate coincidence of the observed $\CC$ and
the matter density, $\ \OL^0\sim \OM^0\,$, i.e.  the ``new'' CC
problem, or ``time coincidence problem'' \cite{weinRMP,CCRev}
{can be alleviated in this framework if we assume the standard
(viz. Appelquist-Carazzone\,\cite{AC}) form of the low-energy
decoupling for the massive fields.}

In the present paper we elaborate on this idea further. We
develop a semiclassical FLRW model whose running CC is driven
smoothly, without fine tuning, due to generic quantum effects
near the Planck scale. We show that, due to the decoupling
phenomenon, the low-energy effects (in particular the physics
from the SM scale) are irrelevant for the CC running, and so the
approximate coincidence between $\OM^0$ and $\OL^0$ is not tied
to any particular epoch in the history of the universe.
Furthermore, the new effects imply deviations from the standard
cosmological equations due to quantum effects. Our
``renormalized'' FLRW model provides a testable framework that
can be thoroughly checked from SNAP data on Type Ia supernovae at
very high z -- see \cite{SNAP, goods} and references in
\cite{rev}. If these experiments  detect a $z$-dependence of the
CC similar to that predicted in our work, we may suspect that
some relevant physics is going on just below the Planck scale.
Alternatively, if they find a static CC, this might imply the
existence of a desert in the particle spectrum near the Planck
scale.

The structure of the paper is as follows. In the next section we
compare constant versus variable CC models. In Section
\ref{sect:RCC} we present our variable CC model based on the
Renormalization Group. In Section \ref{sect:solving} we solve the
FLRW cosmologies with running CC. In Section
\ref{sect:numanalysis} we study the numerical behaviour of these
cosmologies, and the predicted deviations from the standard FLRW
expectations. In Section \ref{sect:newmodel}, we introduce the
magnitude-redshift relation for the analysis of the  SNe Ia. In
Section \ref{sect:snap} we perform the simulations on the SNAP
data in order to test the sensitivity with which the features of
the new model can be determined. In the last section we draw our
conclusions. {Two appendices are included at the end: one to
discuss some technical issues inherent to our QFT framework, the
other providing some background related to the statistical
analysis.}

%%%%%%%%%%%%%%%%%%%%%%%%%%%%%%%%%%%%%%%%%%%%%%%%%%%%%%
\section{Constant versus variable cosmological term}
\label{sect:VCT}

The cosmological constant enters the Hilbert-Einstein (HE) action
as follows
\begin{equation}
S_{HE}=-\int d^{4}x\sqrt{-g}\,\left( \,\frac{1}{16\pi G}%
\,R+\CC _{vac}\,\right) \,.  \label{HE}
\end{equation}
It is well-known that renormalizability requires that this
effective action should be extended with a number of higher
derivative terms \cite{birdav,book}:
\begin{equation}
S_{vac}=\int d^{4}x\sqrt{-g}\,\Big\{
\,a_{1}R^{2}_{\mu\nu\alpha\beta}
+a_{2}R^{2}_{\mu\nu} + a_{3}R^{2} + a_{4}{\square} R-\frac{1}{16\pi G_{vac}}%
\,R - \Lambda_{vac}\Big\}\,.  \label{Svac}
\end{equation}
{The phenomenological impact of the higher derivative terms in
this action is negligible at present, and therefore it suffices
to use the low-energy action (\ref{HE}). However, the presence of
the parameter $\Lambda_{vac}$ is as necessary as any one of these
higher derivative terms to achieve a renormalizable QFT in curved
space-time}\,\footnote{{It follows that quintessence models
without a $\Lambda$ term cannot be renormalizable theories in
curved space-time.}}.

The vacuum CC itself, $\Lambda_{vac}$, is not the physical
(observable) value of the cosmological constant. By definition
the physical CC is the parameter $\Lambda$ entering the
Friedmann-Lema\^{\i}tre equation:
\begin{equation}
H^{2}\equiv \left( \frac{\dot{a}}{a}\right) ^{2}=\frac{8\pi\,G_N }{3}%
\left( \rho +\Lambda\right) -\frac{k}{a^{2}}\,, \label{FL1}
\end{equation}
where $H$ is the expansion parameter (Hubble's ``constant'').
This $\CC$ should be the sum (\ref{lambdaphys}). While the
homogeneous and isotropic FLRW cosmologies do not allow spatial
gradients of $\CC$, they do not forbid the possibility that $\CC$
may be a function of the cosmological time: $\CC =\CC (t)$. In
this case the Einstein field equations associated to the action
(\ref{HE}) read
\begin{equation}
R_{\mu \nu }-\frac{1}{2}g_{\mu \nu }R=-8\pi G_{N}\
\tilde{T}_{\mu\nu}\,, \label{EE}
\end{equation}
where $\tilde{T}_{\mu\nu}$ is given by $\tilde{T}_{\mu\nu}\equiv
T_{\mu\nu}+g_{\mu\nu}\,\CC (t)$, $T_{\mu\nu}$ being the ordinary
energy-momentum tensor associated to matter and radiation. By the
Bianchi identities, it follows that $\CC$ is a constant if and
only if the ordinary energy-momentum tensor is individually
conserved ($\bigtriangledown^{\mu}\,{T}_{\mu\nu}=0$). In
particular, $\CC$ must be a constant if ${T}_{\mu\nu}$ is zero
(e.g. during inflation).

Modeling the expanding universe as a perfect fluid with velocity
$4$-vector field $U^{\mu}$, we have
\begin{equation}
T_{\mu\nu}=-p\,g_{\mu\nu}+(\rho+p)U^{\mu}U^{\nu}\,,
\label{Tmunuideal}
\end{equation}
where $p$ is the isotropic pressure and $\rho$ is the proper
energy density of matter.  Clearly $\tilde{T}_{\mu\nu}$ takes the
same form as (\ref{Tmunuideal}) with $\rho\rightarrow
\tilde{\rho}=\rho+\CC\,,\ \ \ p\rightarrow \tilde{p}=p-\CC$. Using
the FLRW metric
\begin{equation}\label{FLRWm}
  ds^2=dt^2-a^2(t)\left(\frac{dr^2}{1-k\,r^2}
+r^2\,d\theta^2+r^2\,\sin^2\theta\,d\phi^2\right)\,,
\end{equation}
we can compute explicitly the local energy-conservation law
$\bigtriangledown^{\mu}\,\tilde{T}_{\mu\nu}=0$. The result is the
old Bronstein's equation \cite{Bronstein} allowing transfer of
energy between ordinary matter and the dark energy associated to
the $\CC$ term:
\begin{equation}\label{Bronstein}
\dot{\CC}+\dot{\rho}+3\,H\,(\rho+p)=0\,.
\end{equation}
We see that the most general local energy conservation law (or
equation of continuity) involves both the time evolution of
$\rho$ and that of $\CC$. For a truly constant CC, then
$\dot{\Lambda}=0$, and we recover of course the standard
conservation law $\,\,\dot{\rho}+3\,H\,(\rho+p)=0$. Equations
(\ref{FL1}) and (\ref{Bronstein}) constitute two independent
counterparts for constructing FLRW cosmologies with variable
$\CC$. The dynamical equation for the scale factor is
\begin{equation}
\ddot{a}=-\frac{4\pi}{3}G_{N}\,(\tilde{\rho}+3\,\tilde{p})\,a=-\frac
{4\pi}{3}G_{N}\,(\rho+3\,p-2\,\CC)\,a\,, \label{newforce3}
\end{equation}
but it is not independent from the previous two. In the
matter era $p=0$, and Eq. (\ref{newforce3}) shows that unless
$|\CC|$ is much smaller than $\rho$,  a positive $\CC$ eventually
implies accelerated expansion -- as in fact seems to be the case
for our universe \cite{p99,Riess98}.

{It should be clear that our approach based on a variable CC
 departs from all kind of
quintessence-like approaches, in which some slow--rolling scalar
field $\chi$ substitutes for the CC. In these models, the dark
energy is tied to the dynamics of $\chi$, whose phenomenological
equation of state is defined by $p_{\chi}=w\,\rho_{\chi}$. The
term $-2\,\CC$ on the \textit{r.h.s.} of Eq.\,(\ref{newforce3})
must be replaced by
$\rho_{\chi}+3\,{p}_{\chi}=(1+3\omega)\rho_{\chi}$. In order to
get accelerated expansion in an epoch characterized by $p=0$ and
$\rho\rightarrow 0$ in the future, we must require $-w_{-}\leq
w\leq -1/3$, where usually $w_{-}\geq -1$ in order to have a
canonical kinetic term for $\chi$. However, one cannot completely
exclude ``phantom matter-energy'' ($w_{-}< -1$) and
generalizations thereof\,\cite{Stefan}. Present data suggest the
interval $-1.38\leq w\leq -0.82$ at $95\% \,$
C.L.\,\cite{Melchiorri}. Although $p_{\chi}$ and $\rho_{\chi}$ are
related to the energy-momentum tensor of $\chi$, the dynamics of
this field is unknown because the quintessence models do not have
an explanation for the value of the CC. Therefore, the barotropic
index $w$ is not known from first principles. In particular, one
cannot exclude it may have a redshift dependence, which can be
parametrized in various ways as follows:
\begin{equation}\label{wexpansion}
\frac{p_{\chi}}{\rho_{\chi}}\equiv w=w_0+w_1 z +{\cal
O}(z^2)=w_0+w_a (1-a) +{\cal O}((1-a)^2)\,,
\end{equation}
where $z+1=1/a$. Finding a non-vanishing value of $w_1$ ($w_a $)
implies a redshift evolution of the equation of state for the
$\chi$ field \,\cite{WMAP}. The difficulties to
measure $w_1$ are well-known, see e.g.\,\cite{Padmanabhan}.

Quite in contrast to that scenario, since our variable CC is a
``true'' cosmological parameter, the only possible equation of
state for the CC term is $p=-\Lambda$, whether it is a true
constant or it is a parameter that evolves with the cosmological
time. In our case the CC is indeed a variable one, and its
variation is attributed to potential quantum effects linked to
physics near the Planck scale, as will be explained in the next
section.

\section{Renormalization group and cosmological constant}
 \label{sect:RCC}

The possibility of a cosmological model with a time-dependent
$\CC$ as presented in the previous section is very generic.
However, the two differential equations (\ref{FL1}) and
(\ref{Bronstein}) cannot be solved unless a third equation
involving $\CC$ is called for. The third equation admits many
formulations, even at the classical level\,\footnote{See e.g.
\,\cite{ReutWett87a, Overduin} and references therein.}. However,
a particularly interesting implementation occurs when the time
dependence has its prime origin in the quantum field theory
notion of Renormalization Group running \cite{JHEPCC1,cosm}. It
means that the Hilbert-Einstein action (\ref{HE}) is treated
semiclassically and one introduces an equation for the running
cosmological constant. Although this can be done in several ways,
a consistent formulation of the approach has been presented in
\cite{JHEPCC1} within the well established formalism of QFT in
curved space-time (see e.g. \cite{birdav,book}). From simple
dimensional analysis, and also from dynamical features to be
discussed below, the RGE for the physical CC  may take in
principle the generic form \cite{JHEPCC1,Babic}
\begin{eqnarray}
\frac{d\CC}{d\ln\mu}&& =\frac{1}{(4\pi
)^{2}}\left(\sum_{i}\,A_{i}\,m_{i}^{4}
+\mu^{2}\sum_{j}%
\,B_{j}M_{j}^{2}\,\,+\mu^{4}\sum_{j}
\,C_{j}+\mu^{6}\sum_{j}\frac{\,D_{j}}{M_{j}^{2}}\,\,+...\right)\nonumber\\
&&\equiv\,\sum_{n=0}^{\infty}\,\sum_{i}\,\alpha_{in}\,{\cal
M}_{i}^4\,\left(\frac{\mu}{{\cal M}_{i}}\right)^{2n}\,\equiv\bCC
({\cal M}_i,\mu/{\cal M}_i). \label{newRG1}
\end{eqnarray}
where the sums are taken over all massive fields;
$\,A,B,C,D,...\,$ are constant coefficients, and  $\mu$ is the
energy scale associated to the RG running. {We assume that $\mu$
is of the order of some physical energy-momentum scale
characteristic of the cosmological processes, and can be
specified in different ways (see below). In our model we assume
that $\mu$ is given by the typical energy-momentum of the
cosmological gravitons, namely $\mu=H$, which is of order
$R^{1/2}$.} The \textit{r.h.s.} of (\ref{newRG1}) defines the
$\bCC$-function for $\CC$, which is a function of the masses and
in general also of the ratios of the RG scale and the masses.

In the equation above the masses of the various degrees of freedom
(d.o.f.) are represented by $m_i$ and $M_j$. Here we distinguish
between the active (or ``light'') d.o.f. at the scale $\mu$,
namely those satisfying $\mu\gg m_i$ and contributing to the
$\bCC$-function in the form $\sim m_i^4$, from the ``decoupled''
(or ``heavy'') d.o.f. which satisfy  $\mu\ll M_j$ and yield the
remaining terms in the series expansion of $\bCC$ in powers of
$\mu/M_j\ll 1$. As can be seen, all the terms in $\bCC$ are of
the form $\mu^{2n}\,{\cal M}_i^{4-2n}\ (n=0,1,2,3...)$ where
${\cal M}_i=m_i$ for $n=0$ and ${\cal M}_i=M_i$ for $n\geq 1$.
The $n=0$ terms correspond precisely to the active d.o.f.
contributing the full fourth power of their masses. The
coefficients $A_i$ for these terms are known in the ultraviolet
(UV) regime because they must coincide, in any mass-dependent
renormalization framework, with their values in the Minimal
Subtraction (MS) scheme \cite{Collins, LBrown}. {In this regime
the $\bCC$-function depends only on the masses of the active
degrees of freedom, and not on the ratios $\mu/M_i$ of the RG
scale and the heavy masses.} For particles $i=1,2,3,...$ of
masses $m_i$ and spins $J_i$ one finds \cite{JHEPCC1}:
\begin{equation}\label{Ai}
A_i=(-1)^{2J_i}(J_i+1/2)\,n_{J_i}\,N_c\,,
\end{equation}
with $n_{\{0,1,1/2\}}=(1,1,2)$ and $N_c=1,3$ for uncolored and
colored particles respectively. The remaining terms in
(\ref{newRG1}) ``decouple'' progressively faster as we move from
$n=1,2,3,...\ $

Notice that dimensional analysis is not enough to explain the
most general structure of $\bCC$. The fact that only even powers
of $\mu$ are involved stems from the covariance of the effective
action. Indeed, the odd-powers of $\mu$ cannot appear after
integrating out the higher derivative terms, as they must appear
bilinearly in the contractions with the metric tensor. In
particular, covariance forbids the terms of first order in $\mu$.
As a result the expansion must start at the $\mu^2$-order. On the
other hand, the structure for the $n\geq 1$ terms associated to
the coefficients $B$, $C$, ... in (\ref{newRG1}) is dictated
by the the Appelquist-Carazzone (AC) decoupling theorem
 \cite{AC,Collins}. Thus, when applying the AC theorem in
its very standard form to the computation of $\bCC$, the
decoupling does still introduce inverse power suppression by the
heavy masses (those satisfying $M_j\gg\mu$), but since the
$\bCC$-function itself is proportional to the fourth power of
these masses it eventually entails a decoupling law
$1/M^{2n-4}_j$, and so the $n=1$ and $n=2$ terms do not decouple
in the ordinary sense whereas the $n\geq 3$ terms do, i.e. only
the latter start getting (increasingly higher) inverse power
suppression by the heavy masses. The upshot is that, strictly
speaking, the truly decoupling terms in $\bCC$ (in the sense used
when applying the AC theorem to the ordinary SM interactions)
commence at $n=3$ and above. In contrast, the $n=2$ terms are
constant (independent of the masses) and the $n=1$ terms acquire
the peculiar structure $\mu^{2}\,M_{j}^{2}$, hence displaying the
unusual property that a $\beta$-function may increase
quadratically with the heavy masses ({``soft decoupling''}).
Remarkably enough, the CC is the only parameter in the effective
action of vacuum that has the necessary dimension to possess this
distinctive property \cite{JHEPCC1,Babic}, and the latter is
certainly not shared by any other parameter in the SM of the
strong and electroweak interactions.

Despite that the explicit derivation of the decoupling for the CC
is not possible at present (see the extended discussion of this
issue in the Appendix 1), the assumed form of decoupling is
highly plausible \cite{JHEPCC1,Babic} within the general effective
field theory approach\,\cite{Manohar}. Furthermore, the
$\beta_k$-functions for the remaining coefficients
$\,\,a_k\,\,(k=1,2,3,4)$ of the vacuum effective action (viz.
those corresponding to the higher derivative terms in
Eq.\,(\ref{Svac})) do exhibit exactly this kind of decoupling
behaviour assumed for $\,\bCC$\,\cite{apco}. In this situation it
is quite reasonable to apply the phenomenological approach. Since
there are no direct theoretical reasons to exclude the soft
decoupling in the CC sector, we just admit that it really takes
place and investigate the cosmological model which follows from
this assumption. We will see indeed that the $n=1$ structure
(undoubtedly the most peculiar one of the $\bCC$-function) can be
experimentally probed in the next generation of high redshift
cosmological measurements
 \cite{SNAP}-- Cf. Sections \ref{sect:numanalysis}--\ref{sect:snap}.

Following the phenomenological indications, it is very important
that the structure of the $\bCC$-function does not trigger a too
fast running of  $\CC$, which would be incompatible with the
present observations \cite{p99,Riess98}. From
Eq. (\ref{newRG1}) it is clear that this feature will depend not
only on the values of the masses of the various d.o.f. involved,
but also on the characteristic energy scale $\mu$ used to track
the RG running, which must be correctly identified. This is
particularly evident from the quadratic structure of the $n=1$
terms. In the following we mention a few different scenarios that
have been contemplated in the literature:

\begin{itemize}

\item In Ref.\,\cite{cosm} it was assumed that only the
lightest d.o.f. would contribute, equivalently
$B_j=C_j=D_j=...=0\ $ in Eq. (\ref{newRG1}). The only
non-vanishing coefficients here are some of the $A_i\neq 0$,
namely those associated to d.o.f. for which $\mu>m_i$. Typically,
this would be the case for the lightest neutrinos, whose mass can
border the range $m_{\nu}\sim
10^{-3}\,eV$ \cite{AltarelliValle}, and therefore satisfy the
curious numerical coincidence $\CC_0\sim m_{\nu}^4$ which
motivated the RG approach of \cite{cosm}. Moreover, in this
paper the RG scale was identified from the value of the fourth
root of the critical density at a given cosmological time $t$:
\begin{equation}
\,\mu \sim \rc^{1/4}(t)\,, \label{muCD}
\end{equation}
For the present universe, this scale is
$\left(\rc^0\right)^{1/4}\sim 10^{-3}\,eV $, i.e. of the order of
the lightest neutrino mass mentioned above. For the radiation
era, $\rc\sim T^4$ and so $\mu$ in that epoch is essentially
given by the temperature ($\mu\sim T$) within this Ansatz.

\item In Ref. \cite{Babic} the same RG scale (\ref{muCD}) was adopted,
but the important point was made,  on the basis of effective field
theory arguments, to the necessity of including the heavy mass
terms $M_j$ in $\bCC$. Notwithstanding, when applying this
framework to the SM of the strong and electroweak interactions,
where the largest masses are of the order of  a few hundred
$GeV$, one is forced to tame the runaway evolution of $\CC$ --
triggered by the quadratic $n=1$ terms on the \textit{r.h.s.} of
Eq. (\ref{newRG1}). In practice, it means that one has  to
enforce a  fine tuning of their overall effect to
zero\,\cite{Babic},
\begin{equation}\label{finetune}
  \sum_{j}%
\,B_{j}\,\mu^{2}M_{j}^{2}=\left(\rc^0\right)^{1/2}\,\sum_{j}%
\,B_{j}\,M_{j}^{2}=0\,,
\end{equation}
otherwise one gets an extremely fast running of the CC which
would be incompatible with the observations \cite{p99,Riess98}.
The authors of \cite{Babic} use this adjustment to hint at the
value of the Higgs mass $\,M_{H}$, which (for particular values
of the coefficients $B_j$ in their given setting) is the only free
mass parameter in the sum (\ref{finetune}), that runs over all SM
particles. The result that they obtain is reasonable ($M_H\sim
550\,GeV$), but still too high as compared to the current bounds
and expectations \cite{LEPHWG,Hunter}. Moreover, the obtained
value for $M_H$ is scheme-dependent.

\item  Eq. (\ref{newRG1}) was proposed in Ref.\cite{JHEPCC1}
assuming that the RG scale is identified with the square root of
the curvature scalar $\mu \sim R^{1/2}$, which in the FLRW
cosmological context is equivalent to identify $\mu$ with the
expansion parameter (or ``Hubble constant'') at any given
cosmological time:
\begin{equation}
\mu \sim H(t)  \label{muH}\,.
\end{equation}
For the present universe, $\,H_0\sim 10^{-33}\,eV$. This scale is
much smaller than (\ref{muCD}), but from our point of view is the
most natural one, as it is naturally linked with the scale of the
cosmological gravitational quanta (gravitons) -- used here in a
generic sense referring to the presumed quanta of gravity as a
field theory with a tensor potential, rather than to its relation
with the gravitational waves.  Scale (\ref{muH}) is also used
successfully in other frameworks, e.g. in \cite{shocom} to
describe the decoupling of massive particles in anomaly-induced
inflation.

\item  In Ref.\cite{Reuter03a} the RG scale $\mu$ was identified
with the inverse of the age of the universe at a given
cosmological time, i.e. $\mu\sim 1/t$.  This is essentially
equivalent to the previous case, because $H\sim 1/t$ in the FLRW
cosmological setting. Nevertheless, the constitutive relations for
the RG evolution in Ref.\cite{Reuter03a} are different from
\cite{JHEPCC1} and they are phrased in a non-perturbative quantum
gravity framework based on the (hypothetical) existence of an
infrared (IR) fixed point.  In our case, the RG approach aims at
the simplest possible modification of the FLRW cosmology, namely
the study of the CC evolution within perturbative QFT in a curved
background. In contrast to \cite{Reuter03a}, we allow transfer of
energy between matter/radiation and CC, but we do not consider
any significant scaling evolution of Newton's constant. Indeed,
in \cite{JHEPCC1} it was shown that $G_N$ does not undergo any
appreciable running within our perturbative framework.

\item Finally, we consider the framework which we
will elaborate in the rest of this paper. It is based on the
identification
(\ref{muH}) and assumes that the heaviest possible masses
entering Eq. (\ref{newRG1}) lie near the Planck scale, $M_P$
 \cite{Letter,IRGA03}. This approach does not have any
fine-tuning problem in the value of $\bCC$, as we shall see.
\end{itemize}

%%%%%%%%%%%%%%%%%%%%%%%%%%%%%%%%%%%%%%%%%%%%%%%%%%%%%%%%%%%%%%
In the last framework the RGE that supplements (\ref{FL1}) and
(\ref{Bronstein}) is given by a particular form of
Eq.\,(\ref{newRG1}), namely
\begin{equation}
\label{RG1}
\frac{d\CC}{d{\ln\mu}}=\frac{1}{(4\pi )^{2}}\,\sum_i
c_i\,\mu^{2}M_i^{2}+...\,.=\frac{1}{(4\pi )^{2}}\,\sum_i
c_i\,H^{2}M_i^{2}+...\,.
\end{equation}
where $M_i$ is a collection of (superheavy) sub-Planckian-size
masses just below the Planck scale, $M_i^2\lesssim M_P^2$. The
remaining masses are the set of ``low-energy'' masses, $m_i$, in
the sense that $m_i^2\ll M_i^2$,  and therefore do not contribute
in any significant way to this RGE. Since $\mu=H$ is so small at
present, there is not a single d.o.f. satisfying $\mu>m$, i.e.
all coefficients $A_i$ for the $n=0$ terms in Eq.\,(\ref{newRG1})
are zero. Then all the masses are supposed to ``decouple''
according to the soft terms\,\footnote{Let us notice that the
present-day Hubble parameter, $H_0\sim 10^{-33}\,eV$, is 30 orders
of magnitude smaller than the mass of the lightest neutrino, 41
orders of magnitude smaller than the QCD scale and 61 orders of
magnitude smaller than the Planck scale. Obviously, all massive
particles decouple the same way!} ($\sim H^2\,M_i^2 $).

Looking at the decoupling law, the $n=1$ soft decoupling terms are
always the leading ones as compared to all others on the
\textit{r.h.s.} of Eq.\,(\ref{newRG1}). Notice that we assume
$H<M_i$ and that the physics of Planckian or trans-Planckian
energies is governed by some unspecified more fundamental
framework (e.g. string/M theory). Therefore, we do not address
here the issue of whether trans-Plankian physics may also be
responsible for the CC or dark energy in general \cite{TPlanck}.
In fact, we rather propose that the main contribution to the CC
at present can be the sole result of quantum effects from the
highest possible, but still sub-Planckian, energy scales.

We also note that
\begin{equation}\label{numeric}
c\,H_0^{2}M_i^{2}/(4\pi)^2\simeq c\,\left(1.5\
10^{-42}\,GeV\,\times 1.2\
10^{19}\,GeV\right)^2/\left(4\pi\right)^2\sim 10^{-47}\,GeV^4\sim
\CC_0
\end{equation}
for some $c={\cal O}(1-10)$ and $M_i\sim M_P$. The previous result
is very close to the observational data \cite{p99,Riess98}. This
is highly remarkable, because two vastly different and (in
principle) totally unrelated scales are involved to realize this
``coincidence'': $H_0$ (the value of $\mu$ at present) and $M_P$,
being separated by more than $60$ orders of magnitude. In other
words, the ``coincidence'' amounts to saying that the mass scale
associated to the CC at present, $m_{\CC}=\sqrt[4]{\CC_0}\sim
10^{-3}\, eV$, is essentially given by the geometrical mean of the
current value of the Hubble constant and the Planck mass, i.e. the
smallest and largest energy scales conceivable in our universe:
\begin{equation}\label{mCC}
m_{\CC}\simeq\sqrt{H_0\,M_P}\,.
\end{equation}
Eq.\,(\ref{RG1}) provides a possible explanation for that.
Moreover, Eq. \,(\ref{RG1}) tells that the physics of the CC is
naturally dominated by the set of sub-Planckian masses,
irrespective of all the dynamical details of the low-energy
fields with masses $m_i\ll M_i$, such as the SM fields. This idea
completely frees the running of the CC from all kind of
fine-tunings thanks to the smallness of our RG scale $\mu=H$. {In
this suggestive scenario the running of the CC at any time is
smooth enough, in particular also at the present time. At any
epoch the rate of change of the CC is in the right ballpark to
shift the value of CC in less than the value of the CC itself in
that epoch. On the other hand, at higher and higher energies the
RGE (\ref{RG1}) predicts a CC value increasingly larger. For
instance, at the Fermi epoch, when the temperature was of the
order of the Fermi scale $M_F=G_F^{-1/2}\sim 300\,GeV$, the
Hubble parameter was of the order of $H\sim T^2/M_P=M_F^2/M_P$ and
 Eq.\,(\ref{RG1}) predicts a typical value for the CC around $\CC\sim
H^2\,M_P^2\sim M_F^4$, which naturally fits with the value
expected for the CC at the time of the electroweak phase
transition.}

The origin of the Planckian mass operator on the \textit{r.h.s.}
of Eq. (\ref{RG1}) could just be the indelible imprint left
forever on the low-energy physics due to the decoupling of the
sub-Planck mass fields just below the Planck mass scale. This
permanent imprint may be thought of as a ``relic'' low-energy
effect from the high energy dynamics of some fundamental RGE of
the CC at the trans-Planckian scale $\mu>M_i$,
\begin{equation}\label{RGEMP}
\frac{d\CC(\mu)}{d{\ln\mu}}=\frac{1}{(4\pi
)^{2}}\,\sum_i\,A_i\,M_i^4 F_i(\mu/M_i)\,,
\end{equation}
in which the functions $F_i(\mu/M_i)$ of the Planckian masses
$M_i$ depend on the underlying details of the trans-Planckian
physics, e.g. string/M-theory. This Ansatz should hold good
perhaps in the border line $\mu\gtrsim M_P$. Unfortunately we do
not know the details of the RGE (\ref{RGEMP}) as we do not know
the actual structure of the functions $F_i(\mu/M_i)$ for
$\mu>M_i\sim M_P $. Actually for $\mu\gg M_P$ this picture must
break down as it probably does not even make sense to talk of the
Hubble parameter because the metric need not to be the FLRW one.
Indeed, for $\mu > M_P$ we just enter the realm of quantum
gravity, where the metric itself is highly fluctuating. Hence the
Ansatz (\ref{muH}) should be sensible only below the Planck
scale.  Then, and only then, we may set $\mu\simeq H$, and this
identification should be better and better the smaller is the
energy as compared to the Planck scale. It is only at these ``low
energies'' that the soft decoupling of the sub-Planckian masses
dominates the RGE. {For instance, if the form factor takes the
canonical form}
\begin{equation}\label{Fmu}
F(\mu/M_i)\simeq F(H/M_i)=\frac{H^2}{M_i^2+H^2}\,,
\end{equation}
then for $H^2\ll M_i^2$ we may expand the term on the
\textit{r.h.s.} of (\ref{RGEMP}) just to find
\begin{equation}\label{Fmu2}
M_i^4\,\,F(H/M_i)=\,M_i^2\,H^2-\,H^4+ \,\frac{H^6}{M_i^4}+...
\end{equation}
In this way we arrive at some  heuristic justification of
Eq.\,(\ref{newRG1}). At leading order in $\mu$ Eq.\,(\ref{RGEMP})
reduces to our fundamental sub-Planckian operator in (\ref{RG1}).

At super-Planckian energies, $H>M_i$, the form factor
({\ref{Fmu}) is of order one, and the \textit{r.h.s.} of
(\ref{RGEMP}) behaves like $M_i^4$. In this regime we may expect
an RGE of the form
\begin{equation}\label{RGEMP2}
\frac{d\CC(\mu)}{d{\ln\mu}}=\frac{1}{(4\pi
)^{2}}\,\sum_i\,\,A_i\,M_i^4\,.
\end{equation}
Therefore, the typical CC at trans-Planckian energies, just in the
upper neighborhood of $M_P$ ($\mu\gtrsim M_P$), becomes of the
natural size $M_P^4$ and one may ask what to do with it. There is
the attractive possibility that in this Planck neighborhood there
is exact supersymmetry (SUSY), and if so there will be as many
boson fields with mass $M_i$ as fermion fields of the same mass,
and since SUSY applies the sum on the \textit{r.h.s.} of
(\ref{RGEMP2}) could actually vanish identically. Then, at low
energies we find that the CC is always controlled by our leading
term $H^2\,M_P^2$, and when there is a chance for the $M_P^4$
contributions to appear, SUSY kills them automatically. So this
would leave us with a well behaved CC at low energies, and all
the dynamical details associated to the phase transitions below
$M_P$ (in particular the electroweak SM one) would be innocuous
for the running of the CC . This property is robust within the
low-energy regime ($\mu<M_P$) and is guaranteed by the structure
(\ref{RG1}) of the RGE, independent of what particular
speculation is made at $\mu>M_P$ --  e.g. Eq.\, (\ref{RGEMP2})
and the aforementioned SUSY scenario. Of course the SUSY
interpretation is only a possibility and we cannot be too
conclusive. In fact, we cannot say much about the physics at
trans-Planckian energies, not even at the border line $\mu\gtrsim
M_P$, because the relation (\ref{Fmu}) is expected to be valid
only  for $\mu<M_P$. However, this kind of situation is not too
different from what we have with strong interactions in QCD. At
high energies one meets asymptotic freedom, but in the infrared
region the RGE of QCD tells us that the coupling constant grows.
However one can not really conclude that it finally explodes
because we are using an equation that is only valid in the
perturbative regime. Similarly,  here we ignore how the functions
$F(\mu/M_i)$ behave for $\mu>M_P$ and in particular the
identification $\mu=H$ ceases to make sense, so strictly speaking
we cannot use Eq. (\ref{Fmu}) to predict equation (\ref{RGEMP2})
for $H>M_P$. However, like in QCD, we can foresee a plausible
trend in the behaviour for both the low and high energy regimes,
specially if we invoke exact SUSY above $M_P$. Some more
discussion on these issues, including the potential existence of
non-local effects that might appear in the present approach, is
provided in Appendix 1.

\section{FLRW cosmologies with a running cosmological constant}
\label{sect:solving}

\subsection{Solving the model}
\label{sect:DiffEq}

In the previous sections we have motivated our model. Let us now
consider it in detail and show that it is useful and testable.
One has to solve the coupled system of differential equations
formed by Friedmann's equation, the equation of continuity (in the
matter era, where pressure $p=0$) and our RGE, i.e. the system
formed by (\ref{FL1}), (\ref{Bronstein}) and (\ref{RG1}) with
$\mu=H$:
\begin{equation}\label{RG01}
\frac{d\CC}{d{\ln H}}= \frac{1}{(4\pi)^{2}}\
\sigma\,H^{2}M^{2}+...\,.
\end{equation}
Here we have introduced the following mass parameter:
\begin{equation}\label{Mdef}
M\equiv\sqrt{\left|\sum_i\,c_i\,M_i^2\right|}\,.
\end{equation}
Furthermore, $\,\sigma=\pm 1$ indicates the sign of the
overall $\bCC$-function,
depending on whether the fermions ($\sigma=-1$) or bosons
($\sigma=+1$) dominate at the highest energies.  Notice that the
mass $M_i$ of each superheavy particle  in (\ref{RG1}) may be
smaller than $M_P$ and the equality, or even the effective value
$\,M\gtrsim M_P$, can be achieved due to the multiplicities of
these particles. From Eq. (\ref{numeric}) we see that the
\textit{r.h.s.} of (\ref{RG01}) is of the order of the present
value of the CC.

Let us now eliminate the time variable and convert the equation
of continuity (\ref{Bronstein}) into a redshift differential
equation:
\begin{equation}\label{continu2}
\frac{d\Lambda}{dz}\frac{dz}{da}\frac{da}{dt}
\,+\,\frac{d\rho}{dz}\frac{dz}{da}\frac{da}{dt}
\,+\,3\,H\,\rho\,=\,0
\end{equation}
Using the redshift definition $a_0/a=1+z$ and $\dot{a}=a\,H$ it
immediately gives a very simple expression in which both $a$ and
$H$ cancel out:
\begin{equation}\label{continu3}
\frac{d\Lambda}{dz}+\frac{d\rho}{dz}\,=\,\frac{3\,\rho}{1+z}\,.
\end{equation}
One can easily check that if $\Lambda$ would not depend on the
redshift ($d\Lambda/dz=0$) then the previous equation integrates
to $\rho_M\,a^3=\rho_M^0\,a_0^3$, i.e. we recover the old case.
However, in general this is not so and now  Eq. (\ref{continu3})
must be integrated together with (\ref{RG01}) and (\ref{FL1}).
Eq. (\ref{RG01}) can be transformed into a redshift differential
equation by applying again the chain rule:
\begin{equation}\label{RG2}
(4\pi)^{2}\frac{d\CC}{dz}=(4\pi)^{2}\frac{d\CC}{d\ln H}\frac{d\ln
H}{dH}\frac{dH}{dz}=\frac12\,\sigma\,M^{2}\,\frac{dH^2}{dz}\,.
\end{equation}
To compute $dH^2/dz$ we recall Friedmann's equation (\ref{FL1}).
Using the identity $-k/a^2=H_0^2\,\Omega_K^0(1+z)^2$, it takes the
form
\begin{equation}\label{Friedmann2}
H^2(z)=\frac{8\,\pi\,
G}{3}\left[\rho(z)+\Lambda(z)\right]+H_0^2\Omega_K^0\, (1+z)^2\,.
\end{equation}
From this we have
\begin{equation}\label{dHdz}
\frac{dH^2}{dz}=\frac{8\,\pi\,
G}{3}\left(\frac{d\Lambda}{dz}+\frac{d\rho}{dz}\right)+2\,H_0^2\Omega_K^0
(1+z)\,.
\end{equation}
Substituting Eq. (\ref{continu3}) into the previous equation and
then the result into (\ref{RG2}) we find:
\begin{equation}\label{diffrho2}
\frac{d\rho}{dz}\,-\,\frac{3(1-\nu)\,\rho(z)}{1+z}\,
-\,\kappa\,\rc^0\,(1+z)=0\,.
\end{equation}
Here $\rc^0$ is the critical density,
 and we have introduced for convenience two
dimensionless parameters
\begin{equation}\label{bdef}
\kappa \equiv -2\,\nu\OK^0
\end{equation}
and
\begin{equation}
\label{nu}
\nu\equiv \frac{\sigma}{12\,\pi}\,\frac{M^2}{M_P^2}\,.
\end{equation}
Parameter $\kappa$ is related to curvature effects, and it is
not independent of $\nu$ once the spatial curvature $k$ is known.
Our model has one single independent parameter, $\nu$, which will
play an essential role in the forthcoming discussions. The
standard FLRW cosmology corresponds to $\nu=0$. From now on $\nu$
will parametrize all the cosmological functions that we obtain in
our modified (``renormalized'') FLRW framework. For example, the
one-parameter family of solutions of the differential
equation(\ref{diffrho2}) is completely analytical and reads as
follows:
\begin{equation}
\rho(z;\nu) \,=\,\Big(\rM^0+\frac{\kappa}{1-3\nu}\,\rho_c^0\Big)
\,(1+z)^{3(1-\nu)} -\frac{\kappa}{1-3\nu}\,\rc^0\,(1+z)^2\,.
\label{rhoznu}
\end{equation}
The arbitrary constant has been determined by imposing the
condition that at $z=0$ we must have $\rho=\rM^0$. As we have
said, the parameter $\kappa$ above introduces additional
$\nu$-effects due to non-vanishing spatial curvature. If we
assume $\sigma>0$, then  $\kappa>0$ (resp. $\kappa<0$)
corresponds to positively (resp. negatively) curved universes,
i.e. closed (resp. open) cosmologies.  For $\nu=0$ we also have
$\kappa=0$ and one recovers the expected result
$\rho=\rho_M^0\,(1+z)^3$, i.e. $\,\rho\,a^3=\rho_M^0\,a_0^3$.
However, for $\nu\neq 0$ the parameter $\nu$ really plays the role
of a new cosmological ``index'' determining the deviations from
the usual law of evolution with the redshift. Substituting
(\ref{rhoznu}) in (\ref{continu3}) we may explicitly solve also
for the $\nu$-dependent $\CC$, which becomes a function of the
redshift:
\begin{equation}
\CC(z;\nu)= \CC_0+\rM^0\,f(z)\,+\rc^0\,g(z)\,, \label{Lambdaznu}
\end{equation}
with
\begin{eqnarray}
f(z)=\frac{\nu}{1-\nu}\,\left[\left(1+z\right)^{3(1-\nu)}-1\right]\,,
\label{f}
\end{eqnarray}
\begin{eqnarray}
g(z)\,\,=-\frac{\kappa}{1-3\nu}\,\left\{\frac{z\,(z+2)}{2}\,
+\,\frac{\nu}{1-\nu}\,\left[\left(1+z\right)^{3(1-\nu)}-1\right]
\right\}\,. \label{g}
\end{eqnarray}
Notice that the function $f(z)$ is non-vanishing even if the
spatial curvature is zero ($\kappa=0$), whereas $g(z)\neq 0$
introduces curvature effects. To avoid confusion, we note that
$f(z)$ is well defined in the limit $\,\nu\rightarrow 1$.
Similarly, the value $\,\nu=1/3\,$ is non-singular in $\,g(z)$.

We have presented the CC and the matter density function as
explicit functions of the redshift because it is the most useful
way to present the result for astronomy applications. Eq.
(\ref{RG01}) can be trivially integrated with respect to $H$:
\begin{equation}\label{LambdaHz}
\CC(z;\nu)=\CC_0+\frac{\sigma}{2\,(4\pi)^2}\,M^2\,\left(H^2(z;\nu)-H^2_0\right)
=\Lambda_0+\frac{3\,\nu}{8\pi}\,M_P^2\,\left(H^2(z;\nu)-H^2_0\right)\,.
\end{equation}
Computing ${d\CC}/{dz}=({3\,\nu}/{8\pi})\,M_P^2\,{dH^2}/{dz}$ from
this equation, and using (\ref{dHdz}) and the equation of
continuity (\ref{continu3}) to eliminate $d\CC/dz$, it is
immediate to check that we are lead to Eq. (\ref{diffrho2}). This
shows the consistency of the whole procedure.  Not only so;
actually Eq. (\ref{LambdaHz}) can be also useful from the
astronomy point of view because it expresses a relationship
between the CC and the Hubble parameter at any redshift. This
correlation could be an experimental signature of this model,
because it does not take place in the standard model. Furthermore,
since we have already obtained the function $\CC=\CC(z;\nu)$, we
can use it in (\ref{LambdaHz}) to get the explicit function
$H(z;\nu)$. It reads
\begin{eqnarray}
\label{Hzzz}
H^2(z;\nu)&=& H_0^2\,\left\{1+\Omega_M^0\,
\frac{\left(1+z\right)^{3\,(1-\nu)}-1}{1-\nu}  \right.
\nonumber\\
&&\left. +\frac{1-\Omega_M^0-\Omega_{\Lambda}^0}{1-3\,\nu}
\left[(1+z)^2-1-2\nu
\,\frac{\left(1+z\right)^{3\,(1-\nu)}-1}{1-\nu}\right] \right\}\,.
\end{eqnarray}
 It is easy to see that for
$\nu=0$ we recover the standard result: $H^2(z;0)=H^2(z)$, where
\begin{eqnarray}\label{HzSS}
H^2(z)&=& H^2_0\,\left\{1+\Omega_M^0\,\left[(1+z)^3-1\right]+
(1-\Omega_M^0-\Omega_{\Lambda}^0)\,\left[(1+z)^2-1\right]\right\}
\nonumber\\
&=& H^2_0\,\left[\Omega_M^0\,(1+z)^3+
\Omega_{K}^0\,(1+z)^2+\Omega_{\Lambda}^{0}\right]\,,
\end{eqnarray}
and therefore Eq. (\ref{Hzzz}) constitutes a generalization of
this formula for our model. The deviation should perhaps be
testable (see below).

Recall that the evolution in the remote past is obtained in the
limit $z\rightarrow +\infty$ and the asymptotic evolution to the
future corresponds to\,\footnote{This is strictly true only if the
universe expands forever. If not, then it is approximately true,
in the sense that it is valid near the turning point. }
$z\rightarrow -1$. Then, some of the features of $\Lambda(z;\nu)$
and $\rho(z;\nu)$ depending on the value and sign of the
fundamental index $\nu$, are the following:

\begin{itemize}

\item For $\nu<0$ the CC becomes negative and arbitrarily large in the
remote past, and at the same time the matter density infinite and
positive, which is fine provided the latter dominates in the
nucleosynthesis epoch. In the infinite future the CC becomes
finite while the matter density goes to zero. For the flat
universe the finite value of the CC in the asymptotic regime is
positive and given by
\begin{equation}\label{nul1fut2}
  \CC(z=-1)=\CC_0+\left|\frac{\nu}{1-\nu}\right|\,\rM^0\,.
\end{equation}
This case is not incompatible with the measure of a positive CC
in the recent past because all these models satisfy the boundary
condition $\CC (0)=\CC_0$, and therefore if the CC is negative in
the very early universe (anti-de Sitter space) it may just have
changed sign recently. One can easily show that the
transition redshift satisfies
\begin{equation}\label{tz}
\ln(1+z)=\frac{\ln\left[1+(1+\frac{1}
{|\nu|})\,\left({\OL^0}/{\OM^0}\right)\right]}{3\left(1+|\nu|\right)}\,.
\end{equation}
For example, for $\OM^0=0.3$, $\OL^0=0.7$ and $\nu=-0.1$ the
transition from negative to positive $\CC$ occurred around
$z=1.7$. This possibility cannot be excluded in the light of the
present data, which are barely available at such redshifts. For
smaller $|\nu|$, say $\nu=-0.05$, the transition redshift becomes
higher: $z=2.4$. In principle, this $\nu<0$ case is not bad,
except if $|\nu|$ is too large, in which case the transition
redshift would be too low and would have been detected. For
example, for $\nu=-0.4$ the transition  would be at $z=0.69$ (see
Section \ref{sect:numanalysis} for more details). It is thus
clear that the parameter $\nu$ cannot be arbitrary and becomes
restricted by experiment. Moreover, too large $\nu$ (even if
$|\nu|\lesssim 1$) would also lead to problems with
nucleosynthesis (see Section \ref{sect:nucleosynthesis}).

\item Another interesting case is to suppose that
\begin{equation}\label{royal}
0<\nu<1\,,
\end{equation}
where the inequality signs are strict. Then the CC is infinite
and positive in the remote past, and at the same time the matter
density is also infinite and positive. Furthermore, the CC can be
finite in the asymptotic regime while the matter density goes to
zero, which is a double combination of welcome features. The
value of the CC in the future is, in the flat case,
\begin{equation}\label{nul1fut}
  \CC(z=-1)=\CC_0-\frac{\nu}{1-\nu}\,\rM^0\,.
\end{equation}
Of course this is just $\CC_0$ if $\nu=0$. However, for
non-vanishing $\nu$ the CC will be positive or negative in the
asymptotic regime, depending on whether $\,\nu< \OL^0$ or $\,\nu>
\OL^0$ respectively\,\footnote{In the $\CC<0$ case
Eq.\,(\ref{nul1fut}) can only be approximate because the universe
will eventually stop expansion at some $z$ in  $-1<z<0$, see
Eq.\,(\ref{newforce3}).}. The change of sign from $\CC>0$ to
$\CC<0$ in the course of the history of the universe can be of
interest, see below. One can check that this feature could be
maintained in the presence of the curvature term; in particular,
this is so for $\kappa<0$ if $0<\nu<1/3$, and for $\kappa>0$ if
$1/3<\nu<1$.

\item In the flat case, and for $\nu>1$, the CC is finite and positive
in the remote past, where it takes the value:
\begin{equation}\label{nul1past}
  \CC(z=+\infty)=\CC_0+\frac{\nu}{\nu-1}\,\rM^0\,.
\end{equation}
However, this solution seems not to make much sense because the
matter density (\ref{rhoznu}) goes to zero in the remote past.
This would not be the case for $\kappa>0$ because then the
density can go to infinity, due to the second term of
(\ref{rhoznu}).  Notice that the flat case with zero matter and
finite CC in the remote past could momentarily be considered as
tenable in that perhaps the universe was first in a pure state of
CC and then this CC decayed creating matter at much later times.
Such ``decaying CC'' (\ref{nul1past}) could be arbitrarily big if
$\nu\rightarrow 1^+$.  Nonetheless, the $\nu>1$ scenario (with or
without curvature) has a big stumbling block: while the CC
becomes large and negative in the infinite future, the matter
density increases too. This is possible due to the balance of
energy between $\CC$ and $\rho$, Eq. (\ref{Bronstein}), but a
progressively more dense universe looks undesirable because does
not seem to fit the trend of the observed evolution.

\item A critical case is $\nu=1$. Here
the matter density (\ref{rhoznu}) becomes constant for flat space
at all redshifts, and this does not look much sensible. This
scenario, however, could perhaps be rescued for non-vanishing
curvature as follows. For $\kappa>0$ (corresponding to positive
curvature $k>0$ in this case) Eq. (\ref{rhoznu}) tells us that the
matter density is infinite and positive in the remote past and
very small in the long run future, which is what we want.
Furthermore, this situation can be somewhat attractive because it
smoothly matches up with case (\ref{royal}) for $\nu\rightarrow
1^-$. In this limit $f(z)\rightarrow 3\ln(1+z)\,$,
$g(z)\rightarrow (\kappa/2)[z(z+1)/2+3\ln(1+z)]$ and
Eq.\,(\ref{Lambdaznu}) implies that the CC is positive and very
big in the remote past, it eventually changes sign and it starts
getting increasingly negative (remaining finite, though, see the
previous footnote) in the future. It is easy to see that for
sufficiently small (positive) curvature (which is in fact what we
want in order not to depart too much from the flat case), the
transition into the negative CC regime will take place in the
future at the redshift
\begin{equation}\label{ttz}
z\simeq\exp{\left(-\OL^0/3\OM^0\right)}-1\,.
\end{equation}
Thus for $\OM^0=0.30$, $\OL^0=0.71$ (implying $\OK^0=-0.01$) we
get $z=-0.54$, a far point in the future. We stress that although
the $\nu=1$ scenario is not possible in the strict flat space
case, the choice of cosmological parameters that we have made is
perfectly compatible with present day CMB
measurements\,\cite{CMBR,WMAP}. On the other hand the case of
negative curvature would be a disaster because we encounter an
infinitely negative mass density in the remote past.
\end{itemize}

Some reflections are now in order. Take the flat case first. A
most wanted situation for string/M-theory is, as we have
mentioned in the introduction, to have negative (or zero) CC in
the far future enabling to construct the asymptotic S-matrix
states. As we have seen above, a necessary (though not
sufficient) condition for this to happen in the present framework
is to have $\nu>0$. For instance, for $\nu$ in the range
(\ref{royal}) we can start with a large and positive cosmological
constant in the early universe, which then decreases more and
more and eventually, if $\,\nu>\OL^0\,$, it becomes finite and
negative in the asymptotic regime; in fact it can be rather large
and negative if $\nu\rightarrow 1^-$. Actually it suffices that
the CC is negative, no matter how small it is in absolute value,
to secure the eventual stopping of the accelerated expansion and
the disappearance of the event horizon. The largest possible
value of $\nu$ for which  the CC can still change from positive to
negative value is\footnote{The $\nu\rightarrow 1^+$ limit is
troublesome because of the unwanted behaviour of the matter
density in the asymptotic regime. Therefore, $\nu=1$ sets a
barrier and $\nu>1$ should be considered unlikely.} $\nu=1$.
However, this limiting scenario is only tenable at the expense of
having a positively curved universe. In contrast, the
$\OL^0<\nu<1$ solution is in principle allowed in the flat case.
We also remark that there is the possibility to have vanishingly
small CC in the asymptotic future. This would be the case if
$\,\nu= \OL^0$. The values of $\nu$ for which $\Lambda\leq 0$ in
the asymptotic future are smaller than one, but since the present
day estimate of the $\OL^0$ parameter is $\,\OL^0\simeq 0.7$
\cite{p99,Riess98}, the necessary $\nu$ values imply a fairly
large correction to some standard laws of conventional FLRW
cosmology, especially in the flat case\,\footnote{In the curved
cases the solution cannot be obtained in closed form, and we
shall not enter the details here.}. Whether we can accept them or
not is not obvious by now. However, if accepted, then it would
hint at the ``symmetry'' approach to the old CC problem, in the
sense that string/M-theory itself could perhaps provide that
value of $\nu$ as a built-in symmetry requirement. Nonetheless,
before jumping to conclusions, we still have to check what values
of $\nu$ could be incompatible with nucleosynthesis. We do this
in the next section.

%%%%%%%%%%%%%%%%%%%%%%%%%%%%%%%%%%%%%%%%%%%%%%%%%%%%%%%%%%%%%%%%
\subsection{Restrictions from nucleosynthesis}
\label{sect:nucleosynthesis}

Needless to say, it is important to check  what happens with
nucleosynthesis in this model because a non-vanishing $\nu$ may
have an impact not only in the matter-dominated (MD) era, but also
in the radiation-dominated (RD) epoch. We have seen that the index
$\nu$ enters the power of $1+z$ in the expressions for $\rho$ and
$\CC$, and in the MD era we have $\left(1+z\right)^{3\,(1-\nu)}$
rather than the standard behavior $\left(1+z\right)^{3}$.
Similarly, in the radiation era one expects a behaviour of the
form $\left(1+z\right)^{4\,(1-\nu)}$. To check this we recall
that in the RD era the equation of continuity (\ref{Bronstein})
must include the $p\neq 0$ term. For photons the radiation density
$\rho_R$ is related to pressure through $p=(1/3)\,\rho_R$, and we
have
\begin{equation}\label{continu5}
\dot{\CC}+\dot{\rho}+4\,H\,\rR=0\,.
\end{equation}
From the chain rule we can again trade the time variable by the
redshift variable, and the previous formula becomes:
\begin{equation}\label{continu6}
\frac{d\CC}{dz}+\frac{d\rR}{dz}=\rho\,\frac{4}{1+z}\,.
\end{equation}
This equation must now be solved in combination with (\ref{RG01}).
We will not repeat the detailed steps. It is easy to see that the
solution is obtained by simply replacing $3(1-\nu)\rightarrow
4(1-\nu)$ in Eq.\,(\ref{rhoznu})-(\ref{Lambdaznu}).  The radiation
density at any redshift reads:
\begin{equation}\label{rhozR1}
\rR(z;\nu)=\left(\rR^0+\frac{\kappa}{2-4\nu}\right)
\,\left(1+z\right)^{4\,(1-\nu)}-\frac{\kappa}{2-4\nu}\,(1+z)^2\,,
\end{equation}
where
\begin{equation}\label{rhoR}
\rho_R^0\simeq 2.5\times 10^{-5}\,h_{0}^{-2}\,\rho _{c}^{0}
\end{equation}
is the radiation density at present. In this case, since we are in
a thermal bath of radiation, it is more natural to express the
above result in terms of the temperature:
\begin{eqnarray}\label{rhozR2}
\rho_R(T;\nu)&=&\left(\rho_R^0
+\frac{\kappa}{2-4\nu}\right)\,\left(\frac{T}{T_0}\right)^{4\,(1-\nu)}
-\frac{\kappa}{2-4\nu}\,\left(\frac{T}{T_0}\right)^2\nonumber\\
&=& \frac{\pi ^{2}}{30}\,g_{\ast
}\,T^{4}\,\left(\frac{T_0}{T}\right)^{4\,\nu}
+\frac{\kappa}{2-4\nu}\,\left[\left(\frac{T}{T_0}\right)^{4\,(1-\nu)}
-\left(\frac{T}{T_0}\right)^2\right]\,,
\end{eqnarray}
where $T_{0}\simeq 2.75\,K=2.37\times 10^{-4}\,eV$ is the present
CMB temperature.  Of course for $\nu\rightarrow 0$ we recover the
standard result
\begin{equation}\label{rhozphot}
\rho_R(T)=\rho_R^0\,\left(\frac{T}{T_0}\right)^{4}=\frac{\pi
^{2}}{30}\,g_{\ast }\,T^{4}\,,
\end{equation}
with $\,g_{\ast }=2\,$ for photons and $\,g_{\ast }=3.36\,$ if we
take neutrinos into account.  From these equations the
restrictions imposed by nucleosynthesis are rather evident. Let
us first of all quote  the corresponding prediction for the CC in
the radiation epoch, according to this model:
\begin{equation}\label{LambdazR1}
\Lambda_R(T;\nu)=\Lambda_0+\rM^0\,f_R(T)+\rc^0\,g_R(T)\,,
\end{equation}
with
\begin{equation}\label{fzR}
 f_R(T)=\frac{\nu}{1-\nu}
\,\left[\left(\frac{T}{T_0}\right)^{4\,(1-\nu)}-1\right]\,
\end{equation}
and
\begin{equation}
\label{gzR}
g_R(T)\,=\,-\,
\frac{\kappa}{2-4\nu}\,\left\{\frac{T^2-T_0^2}{T_0^2}
\,-\,\frac{\nu}{1-\nu}
\,\left[\left(\frac{T}{T_0}\right)^{4\,(1-\nu)}
-1\right]\right\}\,.
\end{equation}
Again we have separated the result into two functions
$\,f_R(T)\,$ and $\,g_R(T)\,$ in analogy with the MD epoch.
We point out that the limit
$\nu\rightarrow 1/2$ in equations (\ref{rhozR2}) and
(\ref{LambdazR1}) is well defined, as it was also the case with
$\nu=1/3$ in the matter epoch.

In spite of the various possible scenarios that we have described
before for the present MD era, we see that the RD era imposes
additional conditions on the range of values of the cosmological
index $\nu$:

\begin{itemize}

\item $\nu=1$ becomes unfavored by nucleosynthesis.
It is certainly ruled out in the flat case, otherwise the density
of radiation at the nucleosynthesis would be the same as now --
Cf. Eq.\,(\ref{rhozR2}). In the $\kappa> 0$ case the $\nu=1$
scenario could still be argued if one accepts the law $H\propto
T$, instead of $H\propto T^2$, during nucleosynthesis. Actually
this cannot be completely excluded on the basis of existing
phenomenological analyses of the Friedmann equation in the
nucleosynthesis epoch \,\cite{CarrollNucleos}, but in general  we
shall stick here to the most conservative point of view.

\item In the flat case, all $\nu>1$ scenarios are
troublesome because the density of radiation at the
nucleosynthesis time falls below the one at the present time. For
non-vanishing positive curvature, this situation could be
somewhat remedied. However we already noticed that the $\nu>1$
case was untenable because, irrespective of the value of the
curvature, one predicts (in the MD epoch) a progressive growing
of the matter density in the long run future. Therefore, the
$\nu>1$ case remains unfavored.

\item  On the basis of the most conservative set of hypotheses
related to nucleosynthesis we conclude from the structure
(\ref{rhozR2}) of the radiation density, that the safest range for
the $\nu$ parameter is
\begin{equation}\label{royal2}
0\leq\left|\nu\right|\ll 1\,.
\end{equation}
Both signs of $\nu$ are in principle allowed provided the absolute
value satisfies the previous constraint. In the following
we will adhere to this possibility for most of our numerical
analysis, although we shall leave open other possibilities for
the theoretical discussion.

\item Furthermore, when comparing the relative size of the CC,
Eq. (\ref{LambdazR1}), versus the radiation density, Eq.
(\ref{rhozR2}), at the time of the nucleosynthesis we naturally
require that the former is smaller than the latter. For $\kappa=0$
(flat space), the ratio between the CC and the radiation density
is fully determined by the index $\nu$
\begin{equation}
\label{ratioCCrho}
\frac{\Lambda_R(T)}{\rho_R(T)}\simeq \frac{\nu}{1-\nu}\,.
\end{equation}
Then it is clear that in order that the CC is, say, one order of
magnitude smaller than the radiation density at the
nucleosynthesis time, we must again impose Eq. (\ref{royal2}).
Then Eq. (\ref{ratioCCrho}) leads to
\begin{equation}\label{ratioCCrho2}
\frac{\Lambda_R(T)}{\rho_R(T)}\simeq {\nu}\ll 1\,.
\end{equation}

\item  It should be clear that the range (\ref{royal2}) could
already have been suggested from the behaviour of the matter
density function (\ref{rhoznu}) alone, if we are not ready to
tolerate a departure from the exact $\left(1+z\right)^3$ law in
our MD era. However, let us stress that in our era there is no
crucial test (at least an obvious one) emerging from the present
values of the cosmological parameters that is sensitive to the
deviations $\nu\neq 0$ in a way comparable to the nucleosynthesis
test. Maybe the high precision future experiments (see Sections
\ref{sect:numanalysis} and \ref{sect:snap}) can put a remedy to
this. At the moment the restriction on $\nu$ coming from
nucleosynthesis alone coincides with our general will to remain
in the framework of the effective field theory approach
introduced in Section \ref{sect:RCC}. Indeed, Eq. (\ref{royal2})
implies -- see Eq. (\ref{nu})-- that the effective mass scale $M$
cannot be much larger than $M_P$, and in particular $M\lesssim
M_P$ is a natural possibility. This is the kind of picture that
we wanted from the general discussion of our RG framework in
Section \ref{sect:RCC}. So, indeed, the restriction from
nucleosynthesis tells us that we cannot play arbitrarily with the
value of the new cosmological index $\nu$, if we want to get a
consistent picture both theoretically and experimentally.

\end{itemize}

The essential issue is whether the restriction
(\ref{royal2}) leaves still some room for useful phenomenological
considerations at the present matter epoch. The
answer is that it does.  Assume for definiteness that $\sigma=+1$, and let
us take the most natural value for the cosmological index $\nu$:
\begin{equation}\label{nuzero}
\nu_0\equiv\frac{1}{12\,\pi}\simeq  2.6\times 10^{-2}\,,
\end{equation}
which corresponds to $M=M_P$ in (\ref{nu}). Obviously it lies in
the nucleosynthesis suggested range (\ref{royal2}). Let us next
circumscribe the calculation to the flat case, where $\,g(z)=0$.
By expanding the function $f(z)$, Eq. (\ref{f}), in powers of
$\nu$ we immediately find, in first order,
\begin{equation}\label{CCnusmall}
\CC(z)\simeq \CC_0+\nu\,\rho_M^0\,\left[(1+z)^3-1\right]\,.
\end{equation}
What about the numerical effect?  It is given by
\begin{equation}\label{deviLambd}
\delta_{\CC}\equiv \frac{{\CC}(z;\nu)-{\CC}_0}{{\CC}_0}
=\nu\,\frac{\OM^0}{\OL^0}\,\left[(1+z)^3-1\right]\,.
\end{equation}
Taking $\Omega_M^0=0.3$, $\OL^0=0.7$ and $z=1.5$ (reachable by
SNAP \cite{SNAP}) we find $\delta_{\CC}=16.3\%$, namely a
sizeable effect that should be perfectly measurable by SNAP. For
values of order $\nu=0.1$ the previous correction would be as
large as $47.8\%$. This $\nu$ value would correspond to an
effective mass scale (\ref{Mdef}) of $M\lesssim 2\,M_P$.

%%%%%%%%%%%%%%%%%%%%%%%%%%%%%%%%%%%%%%%%%%%%%%%%%%%%%%%%%%%%%%%%
\subsection{The $\nu$-dependent Hubble parameter}
\label{sect:Hparameter}

It is also useful to realize that the value of the Hubble
parameter at a given redshift, is different in our model with
respect to the standard model, see Eq. (\ref{Hzzz}). To check
whether this deviation is testable in the near future, let us see
how much Eq. (\ref{Hzzz}) departs from Eq. (\ref{HzSS}). Since
${\nu}\ll 1$ we can expand Eq. (\ref{Hzzz}) in powers of this
parameter and subtract the standard result (\ref{HzSS}). For
simplicity let us take the flat case ($\kappa=0$). The relevant
subtraction is, in first order of $\nu$,
\begin{equation}\label{subtract}
\Delta (z;\nu)\,\equiv\, H^2(z;\nu)-H^2(z) \,\simeq\, -\,
\nu\,H_0^2\Omega_M^0
\left\{1+(1+z)^3\,\left[3\ln(1+z)-1\right]\right\}\,.
\end{equation}
Therefore, the relative deviation of the Hubble parameter in our
model with respect to the standard one is, at any given redshift
$z$, the following:
\begin{eqnarray}\label{deltaH2}
\delta {H}(z;\nu)&\equiv&
\frac{H(z;\nu)-H(z)}{H(z)}=\frac12\,\frac{\Delta (z;\nu)}{H(z)}
\nonumber\\
&=&-\frac12\,\nu\,\Omega_M^0\,\frac{1+(1+z)^3\,
\left(3\ln(1+z)-1\right)}{1+\Omega_M^0\,\left[(1+z)^3-1\right]}\,,
\end{eqnarray}
where $\,H(z)\,$ is the standard value given by the square root
of (\ref{HzSS}). We see that the correction is negative for
$\,\nu>0$, in which case the model predicts a Hubble constant
smaller than expected at any redshift -- or larger if $\,\nu<0$.
Of course we have $\delta {H}(z=0;\nu)=0$ for any $\nu$, as
expected, because at the present time we have input the values of
the cosmological parameters.  Let us check numbers for the $\nu>0$
case by considering the future SNAP experiment (see Section
\ref{sect:numanalysis} for more details). Take a very distant
supernova at $z=1.7$, and assume that $\nu$ is given by Eq.
(\ref{nuzero}), and that $\OM^0=0.3$. Then Eq. (\ref{deltaH2})
gives $\,\delta {H} \simeq -2.4\%$. The deviation is not big.
Fortunately, the other cosmological parameters are much more
sensitive, and our simulation analyses in Section \ref{sect:snap}
will not be dependent on $H$.  For both signs of $\,\nu\,$ the
Hubble parameter $\,H(z;\nu)\,$ tends to a constant determined by
the cosmological term in the infinite future ($z\rightarrow -1$),
as in the standard case. So we are just testing the different rate
at which $H(z;\nu)$ goes to that constant as compared to
$H(z;0)$. Let us find the asymptotic future regimes for small
$\nu$, even in the presence of curvature. From Eq. (\ref{Hzzz} )
the result is
\begin{eqnarray}\label{asympt}
H^2(z=-1;\nu)&=&
H_0^2\,\left[1-\frac{\OM^0+\OK^0}{1-\nu}\right]\nonumber\\
&\simeq& H^2_0\,\left[\OL^0-\nu\,(\OM^0+\OK^0)\right]
\end{eqnarray}
and for the standard model is of course
\begin{equation}\label{asymptSM}
H^2(z=-1;0)=H^2_0\,\OL^0\,.
\end{equation}
Therefore, in the asymptotic regime the relative difference is,
in first order in $\nu$,
\begin{equation}\label{deltainfty}
\delta {H}(z=-1;\nu)\simeq
-\frac12\,\nu\,\frac{\OM^0+\OK^0}{\OL^0}\,.
\end{equation}
In the flat case ($\OK^0=0$), with $\nu$ as in (\ref{nuzero}) and
the usual values for the cosmological parameters, it gives
$\delta{H}(z=-1;\nu_0)\simeq -0.5\%$.

\subsection{The $\nu$-dependent cosmological sum rule}
\label{sect:sumrule}

From the Friedmann Eq. (\ref{Friedmann2}) we immediately find,
for any $\nu$,
\begin{equation}\label{sumruleshift}
\OM (z;\nu)+\OL (z;\nu)+\OK(z;\nu)=1\,,
\end{equation}
where
\begin{equation}
\OM (z;\nu)\equiv \frac{\rM (z;\nu)}{\rc (z;\nu)}\,,\ \ \
\OL(z;\nu)\equiv \frac{\CC (z;\nu) }{\rc (z;\nu)}\,, \ \ \ \OK
(z;\nu)\equiv \frac{-k }{H^2(z;\nu)\,a^2}\,. \label{cpparameters2}
\end{equation}
Here
\begin{equation}
\rho_{c}(z;\nu) \equiv \frac{3\,H^{2}(z;\nu)}{8\,\pi \,G_{N}}
\label{rhocrit2}
\end{equation}
is the value of the critical density at redshift $z$. Eq.
(\ref{cpparameters2}) reduces to the standard one for $\nu=0$.
The sum rule (\ref{sumruleshift}) is exact at any redshift and
any value of $\nu$ because Hubble's ``constant'' and the critical
density correspond also to that redshift. Now, having said that,
consider the flat case which is well motivated by inflation. The
sum rule is then $\OM (z;\nu)+\OL (z;\nu)=1$, and can (in
principle) be checked experimentally by measuring the Hubble
constant, the matter density and the CC at a given redshift
and then computing the values of the first two parameters in
(\ref{cpparameters2}). However, suppose that we did not suspect
that the CC itself is a function of the redshift and we just
assumed that $\CC=\CC_0$ for all $z$. Then it is clear that the
previous sum rule would fail, because the evolution of the CC and
matter density are indeed related.

Let us consider the explicit expressions for the cosmological
parameters in the flat case. Using the
previous formulae for CC, matter density and Hubble parameter in
our model, we find
\begin{eqnarray}
\OM (z;\nu)&=&\frac{8\,\pi\,G\,\rho_M(z;\nu)}{3\,H^2(z;\nu)}
\,=\,\frac{\OM^0\,\left(1+z\right)^{3\,(1-\nu)}}
{1+\frac{\OM^0}{1-\nu}
\,\left[\left(1+z\right)^{3\,(1-\nu)}-1\right]} \label{OmegaMz}
\end{eqnarray}
and
\begin{eqnarray}
\OL(z;\nu)&=&\frac{8\,\pi\,G\,\CC (z;\nu)}{3\,H^2(z;\nu)}\,=\,
\frac{\OL^0+\frac{\nu}{1-\nu}\,\OM^0\left[(1+z)^{3\,(1-\nu)}-1\right]}
{1+\frac{\OM^0}{1-\nu}\,\left[\left(1+z\right)^{3\,(1-\nu)}-1\right]}\,.\nonumber\\
 \label{OmegaLambdaz}
\end{eqnarray}
The flat space sum rule is clearly borne out at any $z$ and for
all $\nu$,
\begin{equation}\label{SumRuleAllz}
\OM(z;\nu)+\OL(z;\nu)=1\,,
\end{equation}
after making use of the present day sum rule.
  It is also easy
to check the following behaviours in the remote past and in the
infinite future:
\begin{eqnarray}\label{Omegasz}
&& \OM (z=\infty)=1-\nu\,,\ \ \ \ \
\Omega_{\Lambda}(z=\infty)=\nu\,,\nonumber\\
&&  \OM(z=-1)=0\,,\ \ \ \ \ \Omega_{\Lambda}(z=-1)= 1\,.
\end{eqnarray}
Clearly the flat space sum rule is satisfied also in both of
these extreme (past and future) regimes:
\begin{equation}\label{Sumrulenu}
\OM (z=\infty)+\Omega_{\Lambda}(z=\infty)
\,=\,1\,=\, \OM(z=-1)+\Omega_{\Lambda}(z=-1)\,.
\end{equation}
In particular it is interesting to note that in the infinite
future $z\rightarrow -1$ we have
$\Omega_{\Lambda}(z;\nu)\rightarrow 1$ and $\OM
(z;\nu)\rightarrow 0$ both in the standard model ($\nu=0$) and in
our model, irrespective of the value of $\nu$. Indeed, the point
$(\OM=0,\Omega_{\Lambda}=1)$ is a fixed point to which the cosmic
flow is attracted to, and this occurs even for nonzero $\nu$.
However, we remark a difference in behaviour in the remote past:
in the standard case $\OL (z\rightarrow\infty)\rightarrow 0$
whereas in our model it never vanishes and it tends
asymptotically to $\OL (z\rightarrow\infty)\rightarrow \nu$. In
the usual case at hand, with $\nu$ given by Eq. (\ref{nuzero})
and a flat scenario, we find that the $\OL$ parameter in the
remote past was $\OL (z\rightarrow\infty)=0.026$, i.e. some $27$
times smaller than it is now, and it never vanished, contrary to
the standard FLRW model.

With some more algebra one can also check the fulfillment of the
$\nu$-dependent cosmological sum rule in the curved case, for any
$\kappa$. Let us show it for $z=-1$ (the asymptotic regime), as
it can be illustrative. From Eq. (\ref{rhoznu}) it is clear that
$\rho(-1;\nu)=0$ even for $\kappa\neq 0$, so $\Omega(-1;\nu)=0$
too. On the other hand from the complete
expression (\ref{Lambdaznu}) at $z=-1$ (and no expansion in
$\nu$) we have
\begin{equation}\label{lambdam1}
\CC (-1;\nu)=\CC_0-\frac{\nu}{1-\nu}\,\rM^0-
\frac{\nu}{1-\nu}\,(\rho_c^0-\rM^0-\CC_0)\,.
\end{equation}
Moreover, from the $\nu$-dependent Hubble's constant (\ref{Hzzz})
we have
\begin{equation}\label{Hzm1}
H^2(z=-1;\nu)=H_0^2\left[1-\frac{\Omega_M^0+\Omega_{\Lambda}^0}{1-\nu}\right]\,,
\end{equation}
where one could also have used the first line of (\ref{asympt}).
Finally we bring together these two pieces and get
\begin{equation}\label{checkzm1}
\Omega_{\Lambda}(z=-1)
\,=\,\frac{8\,\pi\,G\,\Lambda(z=-1)}{3\,H^2(z=-1;\nu)}\,=\,1\,,
\end{equation}
so the test is successful for all $\nu$ and arbitrary curvature, in the
case $z=-1$. For arbitrary $z$, $\kappa$ and $\nu$, the sum rule
also holds but it is more cumbersome to check.

%%%%%%%%%%%%%%%%%%%%%%%%%%%%%%%%%%%%%%%%%%%%%%%%%%%%%%%%%%%
\subsection{Predicted deviation of $\Omega_{\Lambda}(z;\nu)$
from the standard $\Omega_{\Lambda}(z)$}
\label{sect:deviationOmega}

It should be useful to compute the deviation of the $z$- and
$\nu$-dependent $\Omega_{\Lambda}$ parameter in our framework and
in the standard FLRW case. Let us define the deviation parameter:
\begin{equation}\label{deviOmega}
\delta\Omega_{\Lambda}(z;\nu)=
\frac{\Omega_{\Lambda}(z;\nu)-\Omega_{\Lambda}(z;0)}{\Omega_{\Lambda}(z;0)}\,,
\end{equation}
where $\Omega_{\Lambda}(z;0)=\Omega_{\Lambda}(z)$ is the standard
one at redshift $z$. It is clear that the deviation
(\ref{deviOmega}) must increase more and more with redshift
because the $\nu$-dependent parameter does \textit{not} go to zero
in the past, in contrast to the standard $\Omega_{\Lambda}(z)$,
as shown by equation (\ref{Omegasz}). In the following we
restrict the evaluation of (\ref{deviOmega}) to the flat case.
Substituting Eq. (\ref{OmegaLambdaz}) in (\ref{deviOmega}) and
performing an expansion of the result in powers of $\nu$ we find,
in first order,
\begin{equation}
\label{deviOmegaflat}
\delta\Omega_{\Lambda}(z;\nu)\simeq\nu\,
\left[\frac{{\Omega}_{M}^0\,(1+z)^3-1}{\Omega_{\Lambda}^0}
+\frac{1+3\,{\Omega}_{M}^0\,(1+z)^3\,\ln(1+z)}
{\Omega_{\Lambda}^0+{\Omega}_{M}^0\,(1+z)^3}\right]\,.
\end{equation}
Notice that this formula vanishes exactly at $z=0$,
\begin{equation}\label{checkdev1}
\delta\Omega_{\Lambda}(0;\nu)=0\,,
\end{equation}
as it should because our model and the standard one with the
$\,z$-independent CC have the same initial conditions at $\,z=0$.
Moreover, Eq.\,(\ref{deviOmegaflat}) has the two expected limits
for the remote past and future:
\begin{equation}\label{checkdev2}
\delta\Omega_{\Lambda}(\infty;\nu)=\infty\,,\ \ \ \ \ \ \
\delta\Omega_{\Lambda}(-1;\nu)=0\,.
\end{equation}
In order to visualize  the numerical impact of
(\ref{deviOmegaflat}) in high-z supernovae tests,  let us use the
usual set of inputs: $\nu$ as in (\ref{nuzero}),
$\,\Omega_{M}^0=0.3$, $\,\Omega_{\Lambda}^0=0.7\,$, and take
$\,z=1.5$. The result is
$\delta\Omega_{\Lambda}(1.5;\nu)=20.4\%$. Certainly a $20\%$
correction should be measurable at SNAP. Already for $z=1$ the
result is sizeable: $\delta\Omega_{\Lambda}(1;\nu)=10.2\,\%$. This
$10\%$ effect could be considered as the relative deviation
undergone by $\Omega_{\Lambda}$ when we compare the known set of
high redshift supernovae used in this paper, whose average
redshift is around $z=0.5$ (and from which the parameters
$\Omega_{M}^0=0.3$, $\Omega_{\Lambda}^0=0.7$ were determined),
with the central value $z=1.5$ of the highest redshift set to be
used by SNAP.

%%%%%%%%%%%%%%%%%%%%%%%%%%%%%%%%%%%%%%%%%%%%%%%%%%%%%%%%%%%
\subsection{Deceleration parameter}
\label{sect:deceleration}

It is also interesting to look at the deviations of the
deceleration parameter with respect to the standard model. It is
well-known that there are already some data on Type Ia supernovae
located very near the critical redshift $z^{*}$ where the
universe changed from deceleration to acceleration
\cite{transition}. But of course the precise location of $z^{*}$
depends on the FLRW model and variations thereof. In our case
$\,z^*\,$ should depend on our cosmological index $\nu$, i.e.
$z^{*}=z^{*}(\nu)$. The definition of deceleration parameter
leads to
\begin{equation}
\label{pa1}
q(z;\nu)=-\frac{\ddot{a}\,a}{\dot{a}^2}
=-\frac{\ddot{a}}{a\,H^2(z;\nu)}=\frac12\,\left[\OM
(z;\nu)-2\,\OL (z;\nu) \right]\,.
\end{equation}
Equivalently,
\begin{equation}\label{pa2}
q(z;\nu)\,=\,-1-\frac{\dot{H}}{H^2}
\,=\,-1+\frac12\,(1+z)\,\frac{1}{H^2(z;\nu)}\,\frac{dH^2(z;\nu)}{dz}\,.
\end{equation}
For simplicity in the presentation, let us consider the flat
case. Substituting either Eq. (\ref{OmegaMz}) and
(\ref{OmegaLambdaz}) in (\ref{pa1}), or just Eq. (\ref{Hzzz}) in
(\ref{pa2}), and expanding in first order of $\nu$ we find the
$\nu$-dependent deceleration parameter,
\begin{eqnarray}\label{panu}
&&q(z;\nu)=-1+\frac{3}{2}\,\frac{\OM^0\,(1+z)^3}{1+\OM^0\left[(1+z)^3-1\right]}\\
&&\times\left\{1-\nu\left[3\ln(1+z)+
\frac{\OM^0\left[(1+z)^3-1-3(1+z)^3\ln(1+z)\right]}
{1+\OM^0\left[(1+z)^3-1\right]}\right]\right\}\nonumber\,.
\end{eqnarray}
For $\,z=0\,$ we recover the standard FLRW
result
\begin{equation}\label{pa22}
q(0;\nu)=-1+\frac{3}{2}\,\OM^0=\frac{\OM^0}{2}-\,\OL^0\equiv q_0
\end{equation}
for a flat universe. Of course the last formula is independent of
$\,\nu$, because we have normalized our inputs to reproduce the
cosmological parameters at present. On the other hand for
$\nu=0$, but any $z$,
\begin{eqnarray}
\label{panu2}
q(z;0)\,=\,-1+\frac{3}{2}\,
\frac{\Omega_M^0\,(1+z)^3}{1+\Omega_M^0\left[(1+z)^3-1\right]}\,.
\end{eqnarray}
This is the deceleration parameter as a function of the redshift
for a standard flat FLRW universe. This result vanishes at the
redshift $z=z^{*}$, where
\begin{equation}\label{zstar}
z^{*}=-1+\sqrt[3]{2\,\frac{\OL^0}{\OM^0}}\simeq 0.67\,,
\end{equation}
for $\OL^0=0.7$ and $\OM^0=0.3$. Hence the value (\ref{zstar})
represents the transition point from a decelerated regime
$q(z;\nu)>0$ (corresponding to $z>z^{*}$), into an accelerated
one $q(z;\nu)<0$ (corresponding to $z<z^{*}$), within the flat
FLRW standard model. It should be clear that this transition, for
any curvature, does \textit{not} represent the border crossing
from a matter dominated universe, $\rM (z)>\,\CC (z)$, into a CC
dominated universe, $\rM (z)<\,\CC (z)$ (actually this crossing
occurs later at $z\simeq 0.33$); it rather represents (see Eq.
(\ref{pa1})) the transition from the era where $\OM (z)>2\,\OL
(z)$ (decelerated expansion) into the era where $\OM (z)<2\,\OL
(z)$ (accelerated expansion) -- therefore from $\rM (z)>2\,\CC
(z)$ to $\rM (z)<2\,\CC (z)$. For $\nu\neq 0$, and in the
particular case of the flat space, it defines the transition from
$\OL (z;\nu)<1/3$ to $\OL (z;\nu)>1/3$. Then it is not difficult
to see that the following inequality defines the value of
$\,z^{*}(\nu)$:
\begin{equation}\label{inflexion}
\frac{1}{2}\,(1+z)^3-\frac{\Omega^0_{\Lambda}}{\Omega^0_M}\,\,<\,\,
\nu\,\left\{(1+z)^3\left[1+\frac{3}{2}\,\ln(1+z)\right]-1\right\}\,.
\end{equation}
If this inequality is satisfied, it means acceleration; if it is
violated it means deceleration. The $\nu$-dependent transition
point $z^{*}(\nu)$ is defined by the equality of both sides.
Notice that for $\nu=0$ it immediately reproduces the previous
result (\ref{zstar}). The inequality cannot be solved analytically
for $\nu\neq 0$, but in the next section we provide the numerical
results. Already analytically it is obvious that for $\nu<0$  the
critical redshift $z^{*}(\nu)$ will become smaller (closer to our
time) than (\ref{zstar}), whereas for $\nu>0$ the value of
$z^{*}$ will be larger, i.e. the transition from deceleration to
acceleration occurs earlier. In the presence of curvature
($\kappa\neq 0$) the analytical expressions defining the
transition point are more cumbersome and we limit ourselves to
present the numerical results in Section \ref{sect:numanalysis}.

\subsection{A note on inflation and the RG approach}
\label{sect:inflation}

As a special theoretical issue concerning the RG framework
presented here, let us say a few words on how to potentially
incorporate inflation. Our $\nu$-dependent cosmological equations
(Cf. Sections \ref{sect:DiffEq} and \ref{sect:nucleosynthesis})
predict, for $0<\nu\ll 1$, that at higher and higher energies
there is a simultaneous increase of both the matter/radiation
energy density and CC term. For example, at some Grand Unified
Theory (GUT) scale $M_X$, where $H\sim M_X^2/M_P$, our RGE
(\ref{RG01}) naturally predicts $\CC\sim M_X^4$. During the RD
epoch the CC is always smaller than the radiation energy density
by a factor $\nu\ll 1$, Eq.\,(\ref{ratioCCrho2}). However, this
picture must break down at the very early epoch where radiation
is not yet present (matter-radiation are still to be ``created'').
It is conceivable that this fast inflation period occurs near the
Planck scale, following e.g. the anomaly-induced mechanism
suggested in \cite{shocom}, which is a modification of the
original Starobinsky's model \cite{star}. If there is SUSY, as
speculated in Section \ref{sect:RCC}, and this symmetry breaks
down at some energy near $M_P$, then there is no contribution to
the CC running above that energy. However, just below $M_P$ a CC
of order $M_P^4$ is induced (Cf. Eq. (\ref{RGEMP2})) due to the
mismatch between the boson and fermion masses at that scale, and
so inflation can proceed very fast. As inflation evolves
exponentially the scale $\mu=H$ decreases and the SUSY particles
decouple progressively. Since the total number of scalar and
fermion d.o.f. lessens with respect to the number of vector boson
d.o.f., the anomaly-induced inflation mechanism becomes unstable
and it finally leads to a FLRW phase-- see the details in
Ref.\cite{shocom}. At this point the RD epoch of the FLRW
universe starts: the radiation is supposed to have emerged from
the decaying of that vacuum energy density, of order $M_P^4$. Of
course we have $H<M_P$ after inflation, and the RGE has already
changed to Eq. (\ref{RG01}). So, following the above discussion,
we are left again with the ratio (\ref{ratioCCrho2}), which
insures $\CC_R\ll \rR$ and hence safe nucleosynthesis. The
details of the combined mechanisms will not be discussed here,
but it is clear that the model of Ref. \cite{shocom} can be
naturally invoked in our CC approach because that model is based
on the decoupling of the heavy degrees of freedom according to
the RG scale (\ref{muH}), exactly as in the present framework.

%%%%%%%%%%%%%%%%%%%%%%%%%%%%%%%%%%%%%%%%%%%%%%%%%%%%%%%%%%%%%%%%
\section{Numerical analysis of the model}
\label{sect:numanalysis}

In order to see the behavior of the most representative parameters
describing the universe, we analyze numerically the results
obtained in Section \ref{sect:solving} for the physically
interesting values of the cosmological index, $|\nu| \ll 1$. In
particular we use $\nu=0,\,\pm\nu_0,\,\pm 2\nu_0, \,\pm 0.1$,
where $\nu_0$ is given by (\ref{nuzero}).

We will first of all concentrate on the flat case and later on we
consider an extension to $\,k=\pm 1\,$. In the following we take
$\,\Omega_M^0=0.3\,$ and $\,\Omega_{\Lambda}^0=0.7\,$ at $z=0$
for a flat universe, $\,\Omega_M^0=0.3$,
$\,\,\Omega_{\Lambda}^0=0.5\,$ for an open universe, and
$\,\Omega_M^0=0.4$, $\,\,\Omega_{\Lambda}^0=0.8\,$ for the closed
case. Spatially curved universes are not favored nowadays by  CMB
data\,\cite{CMBR}, but we would like, nevertheless,
 to show their behaviour for some cases which significantly
deviate from flatness. The kind of study we present here is mainly
based on supernova data, and we treat this analysis independently
from CMB measurements. Implications of the RG framework for the
CMB will not be discussed here.

%%%%%%%%%%%%%%%%%%%%%%%%%%%%%%%%%%%%%%%%%%%%%%%%%%%%%%%%%%%%%%%%
\subsection{Flat universe}

Let us start with an universe with flat spatial section. In this
case the evolution of the matter density and of the CC is shown
in Fig.\,\ref{fig:evmupla1}a,b. These graphics illustrate
Eq.\,(\ref{rhoznu}) and (\ref{Lambdaznu}) for $k=0$. As a result
of allowing a non-vanishing $\bCC$-function for the CC
(equivalently, $\nu\neq 0$) there is a simultaneous, correlated
variation of the CC and of the matter density. The evolution of
$\rM (z;\nu)$ and $\CC (z;\nu)$ with $z$ and $\nu$ is very
relevant because these functions appear directly in the luminosity
distance expression -- Cf. Section \ref{sect:newmodel}.

Comparing with the standard model case $\,\nu=0\,$ (see the
Fig.\,\ref{fig:evmupla1}a,b), we see that for a negative
cosmological index $\,\nu\,$ the matter density grows faster
towards the past ($z\rightarrow\infty$) while for a positive
value of $\,\nu\,$ the growing is slower than the usual
$\,(1+z)^3$. Looking towards the future ($z\rightarrow -1$), the
distinction is not appreciable because for all $\nu$ the matter
density goes to zero. The opposite result is found for the CC,
since then it is for positive $\nu$ that $\CC (z;\nu)$ grows in
the past, whereas in the future it has a different behaviour,
tending to different (finite) values in the cases $\,\nu<0\,$ and
$\,0<\nu<1$, while it becomes $\,\,-\infty\,$ for $\,\nu\geq 1$
(not shown). We have made some general comments on these
behaviours in Section \ref{sect:solving}, and given the limiting
formulas for these cases; here we just display some exact
numerical evolutions with $\,z\,$ within the relevant intervals.

\figone

In the phenomenologically most interesting case $\,|\nu|<1\,$ (see
Section \ref{sect:nucleosynthesis}) we always have a null density
of matter and a finite (positive) CC in the long term future,
while for the far past yields $\,\CC=\pm\infty\,$ depending on
the sign of $\,\nu$.  In all these situations the matter density
safely tends to $\,+\infty$. One may worry whether having
infinitely large CC and matter density in the past may pose a
problem to structure formation. From Fig.\,\ref{fig:evmupla1}a,b
it is clear that there should not be a problem at all since in our
model the CC remains always smaller than the matter density in
the far past, and in the radiation epoch $z>1000$ we reach the
safe limit (\ref{ratioCCrho2})}. Actually the time where $\CC
(z;\nu)$ and $\rM (z;\nu)$ become similar is very recent. Take,
for example, the flat case and assume the usual values of the
cosmological parameters as in Fig.\,\,\ref{fig:evmupla1}: then for
$\nu=(-2\nu_0,-\nu_0,\, 0\,, +\nu_0,+2\nu_0)$ equality of CC and
matter density takes place at $z=(0.29, 0.31, 0.32, 0.34, 0.36)$
respectively. For larger values of $\nu$ (still in the $|\nu|\ll
1$ range), say $\nu=(-0.1,+0.1)$, we find  $z=(0.27, 0.43)$. In
all cases the equality of matter density and CC corresponds to
very recent times, and so the evolution of the CC in this model
never prevented structure formation.

%%%%%%%%%%%%%%%%%%%%%%%%%%%%%%%%%%%%%%%%%%%%%%%%%%%%%%%%%%%%%%%%%%%%%%%%%%%%
\figtwo
%%%%%%%%%%%%%%%%%%%%%%%%%%%%%%%%%%%%%%%%%%%%%%%%%%%%%%%%%%%%%%%%%%%%%%%%%%%%

{Related to the evolution of $\rM (z;\nu)$ and $\CC (z;\nu)$ are
the cosmological parameters $\OM (z;\nu)$ and $\OL (z;\nu)$
respectively -- see Fig.\,\ref{fig:OMOL}a,b. They are sensitive to
the evolution of the corresponding energy densities and at the
same time to that of the Hubble expansion rate $H(z;\nu)$ -- see
Eq. (\ref{Hzzz}). The evolution of  $\OM (z;\nu)$ and $\OL
(z;\nu)$ tells us how the present values $\OM^0$ and $\OL^0$
differ from the corresponding values in the past and in the
future for the standard model case ($\nu=0$) and the present
model case ($\nu\neq 0$). Although $\OL$ is found to be at
present of the order of $70\%$ of the matter/energy in the
universe, as we approach $z\sim1$ its contribution was only a
quarter of the total. The exact value obviously depends on the
value of $\nu$ (Fig.\,\ref{fig:OMOL}b), being already  $\sim 10\%$
larger/smaller than for a constant $\CC$ for $\nu=\pm \nu_0$ at
redshift $z=1$. If we go further back in time, $\OL (z\rightarrow
+\infty)$ always diminishes, and tends asymptotically to $\nu$,
while in the standard case $\,\OL (z\rightarrow\infty)=0$. Just
the opposite occurs for  $\OM (z;\nu)$ (Fig.\,\ref{fig:OMOL}a),
since we are showing the flat case and so the sum has to be 1.

%%%%%%%%%%%%%%%%%%%%%%%%%%%%%%%%%%%%%%%%%%%%%%%%%%%%%%%%%%%%%%%%%%%%%%%%%%%%
\figthree
%%%%%%%%%%%%%%%%%%%%%%%%%%%%%%%%%%%%%%%%%%%%%%%%%%%%%%%%%%%%%%%%%%%%%%%%%%%%
%%%%%%%%%%%%%%%%%%%%%%%%%%%%%%%%%%%%%%%%%%%%%%%%%%%%%%%%%%%%%%%%%%%%%%%%%%%%
\figfour
%%%%%%%%%%%%%%%%%%%%%%%%%%%%%%%%%%%%%%%%%%%%%%%%%%%%%%%%%%%%%%%%%%%%%%%%%%%%
%%%%%%%%%%%%%%%%%%%%%%%%%%%%%%%%%%%%%%%%%%%%%%%%%%%%%%%%%%%%%%%%%%%%
\figfive

%%%%%%%%%%%%%%%%%%%%%%%%%%%%%%%%%%%%%%%%%%%%%%%%%%%%%%%%%%%%%%%%%%%%
%%%%%%%%%%%%%%%%%%%%%%%%%%%%%%%%%%%%%%%%%%%%%%%%%%%%%%%%%%%%%%%%%%%
\figsix

%%%%%%%%%%%%%%%%%%%%%%%%%%%%%%%%%%%%%%%%%%%%%%%%%%%%%%%%%%%%%%%%%%%%

In Fig.\,\ref{fig:evmupla2}a,b,c we show the three deviations of
the parameters $\delta \CC$, $\delta \OL$ and $\delta H$ with
respect to the standard model case ($\nu=0$), as defined in
Section \ref{sect:nucleosynthesis}, \ref{sect:deviationOmega} and
 \ref{sect:Hparameter} respectively.
Consider first the deviation of $\,\CC (z;\nu)$ caused by the
running. As a function of $\,z\,$ and $\,\nu$, $\,\CC (z;\nu)\,$
is very sensitive to $\,\nu\,$ at large $\,z$. Thus at $\,z\sim
2\,$ the increment of $\,\CC$ is of $50\%$ for $\,\nu=2\nu_0$,
and once again the effect is higher for negative $\nu$
(Fig.\,\ref{fig:evmupla2}a). Similar or even larger numbers are
obtained for $\delta \OL$, which attains e.g. $70\%$ under the
same conditions. In contrast, the deviations $\delta H$ of the
Hubble parameter from the standard value are much smaller
(Fig.\,\ref{fig:evmupla2}c), around $6\%$. Finally, the CC
variation rate with redshift (normalized to the current critical
density), $(1/\rc^0)\,d\CC/dz$, is presented in
Fig.\,\ref{fig:evmupla2}d. These curves are non-symmetric in the
sign of $\nu$ when $z$ becomes large. The effect is more
important for negative $\nu$. As one can already see from $\CC
(z;\nu)$, Eq. (\ref{Lambdaznu}), as we go further in redshift the
running increases more quickly for $\nu<0$ than for $\nu>0$.

To conclude  the analysis of the flat case we consider another
relevant exponent describing how the universe evolves, the
deceleration parameter {\it q}. This one is fully sensitive to
the kind of high-z SNe Ia data under consideration. The
$\,\nu$-dependence of this parameter was discussed in detail in
Section \ref{sect:deceleration}. The transition point between
accelerated and decelerated expansion is a function of $\,\nu\,$:
the more negative is $\,\nu$, the more delayed is the transition
(closer to  our time)-- see Fig.\,\ref{fig:qHpla}. If $\,\nu>0$,
the transition occurs earlier (i.e. at larger $z$). While in the
standard case, and for a flat universe, the transition takes
place at redshift 0.67 --Eq. (\ref{zstar}) --, it would have
occurred at $z=0.72$ and $z=0.78$ for $\nu=\nu_0$ and
$\nu=2\nu_0$ respectively, and at $z=0.63$ and $z=0.59$ for
$\nu=-\nu_0$ and $\nu=-2\nu_0$ (Cf. Fig.\,\ref{fig:qHpla}). For
$\nu=(-0.1,+0.1)$ the effect is quite large, namely the
transition would be at $z=(0.53,0.91)$ and hence there is a
correction of $(-21\%,+36\%)$ with respect to the standard case.

%\clearpage

%%%%%%%%%%%%%%%%%%%%%%%%%%%%%%%%%%%%%%%%%%%%%%%%%%%%%%%%%%%%%%%%%%%%%%%%%%%
\subsection{Curved universe}

For the small values of $\,\nu\,$ that we present in the previous
figures, the differences between a flat $\,k=0\,$ universe and a
$\,k=\pm 1\,$ one are not evident. A trivial variation is that
coming from the different choice of the present-day values of
$\,\Omega_M^0\,$ and $\,\Omega_{\Lambda}^0$. Besides, we have
more marked $\nu$--dependent features than in a flat universe,
but the main characteristics remain the same.  In order to see the
differences for the curved case, we show in Figures
\ref{fig:derivada_cur} and \ref{fig:curv}
 the most representative parameters:
$\,\rho_M, \,\Lambda, \,\Omega_{\Lambda}\,$ and
$\,{d\Lambda}/{dz}\,$, both for positive and negative curvature.
Qualitatively, the behaviours are similar
 to the flat case. However we note that it is for positive curvature, i.e. closed
universe, that we find the most dramatic numerical differences
with respect to the flat case for each value of the cosmological
index $\nu$. In particular, for closed universes we observe a
faster running of the CC, especially for $\nu<0$ (this fact is
confirmed when adopting other sets of cosmological parameters for
closed universes). Therefore, $k=+1$ universes would represent
the most favored case for the possibility of the observational
detection of the CC running in our model. On the other hand the
open universes differ numerically very slightly from the flat case
even though the degree of (positive or negative) curvature chosen
in the two examples shown in Figures \ref{fig:derivada_cur} and
\ref{fig:curv} is the same, namely $\OK=\pm 0.2$. As already
advertised, from the point of view of the CMB data the curved
cases under consideration would be excluded because $\OZ=1.02\pm
0.02$\,\cite{WMAP}. However, we should still be open minded to
the possibility of non-flat universes and maintain full
independence of the two sets of data, CMB and high redshift SN
Ia.  Although we do not display the behavior of the quantities
presented in the previous section due to their similarities, it
might be interesting to comment where the transition from
deceleration to acceleration takes place. In the case of the open
and closed universes defined above the transition redshifts for
$\nu=(-0.1 -\nu_0, -2\nu_0,  0, \nu_0 , 2\nu_0, 0.1)$ read:
$z=(0.36, 0.42, 0.45, 0.49, 0.55, 0.61, 0.77)$ for open, and
$z=(0.49, 0.53, 0.56, 0.59, 0.62, 0.65, 0.73)$ for closed. So,
the width of the redshift interval is almost the same as for the
flat universe, but the transitions tend to occur at lower
(higher) redshifts for open (closed) universes.

\clearpage
\section{Magnitude-redshift relation and Type Ia Supernovae}
\label{sect:newmodel}

The analysis of supernova magnitudes allows to test different
cosmological models since magnitude depends on the dynamical
evolution of the universe. This is true not only for supernovae
but for all standard candles, i.e., objects whose absolute
magnitude $M$ (or intrinsic luminosity ${\cal{L}}$) we know, and
whose apparent magnitude $m$ (or received flux ${\cal{F}}$) can be
measured at a given redshift. For cosmological purposes related
to the expansion history of the universe, SNe Ia are the best
candles since their luminosity enables detection up to very high
redshift. Besides, although there are no standard candles in
nature, in the SNe Ia case the variety is well accounted for by a
tight correlation between magnitude at maximum and  decline of the
light curve
 \cite{p99,rev}, and the candle can be calibrated through such correlation.
A way to parameterize this effect is through the stretch factor,
{\it s}, used by the Supernova Cosmology Project (SCP) \cite{p99}.
The stretch factor method expands or contracts by a factor $s$ the
time axis of every supernova light curve to fit a fiducial one.
The {\it stretch--corrected} SNe Ia magnitudes of the sample are
then fitted  to obtain the cosmological parameters.

The apparent magnitude obtained at different redshifts is related
to a given cosmological model via the magnitude-redshift relation. One
starts from the notion of luminosity distance, $d_L$, related to
the received flux ${\cal{F}}$ and the absolute (intrinsic)
luminosity ${\cal{L}}$ through the geometric
definition \cite{Peebles}:
\begin{equation}\label{flux}
{\cal{F}}=\frac{{\cal{L}}}{4\pi d_L^2}\,.
\end{equation}
Then the logarithmic relation between flux and (theoretical)
apparent magnitude reads
\begin{equation}\label{magnitud}
m^{th}(z,H_0,\OM^0,\OL^0) \,=\, {\cal{M}}\,+\,
 5\log_{10}\left[H_0\, d_L(z,H_0,\OM^0,\OL^0)\right]\,.
\end{equation}
In the last equation, terms have been defined in order to collect
all the dependence on the current value of the Hubble parameter
$H_0$ into the expression
\begin{equation}\label{m0}
{\cal{M}} = M - 5\log_{10} H_0 \,+\, 25\,.
\end{equation}
This way all the model dependence is encoded in the
luminosity-distance function $d_L=d_L(z,H_0,\OM^0,\OL^0)$. Notice
that the combined expression $H_0\,d_L$ entering the argument of
the log on the \textit{r.h.s.} of Eq. (\ref{magnitud}) is Hubble
constant-free. On the other hand, from the FLRW metric
(\ref{FLRWm}), the luminosity distance of a source at
(dimensionless) radial coordinate $r$ and redshift $z$ is given
by $d_L=a_0\,\, (1+z)\,r$, where $a_0=a(t_0)$. So we need to
compute $r$ as a function of the cosmological parameters. Since
our FLRW universes have been modified by the renormalization
group effects represented by the $\nu$-parameter, the
luminosity-distance relation takes a slightly different form as
compared to the standard one \cite{Peebles}. This can be foreseen
from the generalized structure of the $\nu$-dependent expansion
rate (\ref{Hzzz}) or, equivalently, the $\nu$-dependent
cosmological constant parameters (\ref{cpparameters2}). It means
that in our modified FLRW model the luminosity distance becomes a
function of $\,z\,$ parameterized by $\,\nu\,$ and the present
day values of the cosmological parameters:
$\,\,d_L=d_L(z,H_0,\OM^0,\OL^0;\,\nu)$. For $\nu=0$ this function
reproduces the standard result. The explicit derivation of this
function for $\nu\neq 0$ follows steps similar to the conventional
case, namely one starts considering the equation
$a\,dr=dt\,\sqrt{1-k\,r^2}$ for a null geodesic along a radial
direction, which follows from the FLRW metric (\ref{FLRWm}). This
can be rewritten as
\begin{equation}
\label{pas2}
\frac{dr}{\sqrt{1-k\,r^2}}=\frac{1+z}{a_0}\,dt
=-\frac{1}{a_0}\,\frac{dz}{H(z)}\,,
\end{equation}
where $H(z)$ is the expansion rate at redshift $z$. Upon
integration on both sides we have
\begin{equation}
\label{pas3}
\int_{0}^{r}\,\frac{dr'}{\sqrt{1-k\,{r'}^2}}
=\frac{1}{a_0}\,\int_{0}^{z}
\,\frac{dz'}{H(z',\OM^0,\OL^0;\nu)}\,,
\end{equation}
where we have taken into account that in our case the expansion
rate $H(z,\OM^0,\OL^0;\nu)$ is the $\nu$-dependent function
$H(z;\nu)$ given by Eq. (\ref{Hzzz}). After trivial integration of
the \textit{l.h.s.} of Eq. (\ref{pas3}) for $k=0,\,\pm 1$ one
finds the desired radial function $r=r(z,k,H_0,\OM^0\,\OL^0)$.
Trading the curvature parameter $k$ for $\OK^0=-k/H_0^2\,a_0^2$,
one immediately finds the exact luminosity-distance function
\begin{equation}\label{dist}
d_L(z,H_0,\OM^0,\OL^0;\nu) = \frac{1+z}{H_0 \sqrt{|\OK^0|}}\ \Psi
\left(
 \sqrt{|\OK^0|} \int_{0}^{z}{\frac{H_0 ~ dz'}{H(z',\OM^0,\OL^0;\nu)}}
 \right)\,,
\end{equation}
with
\begin{equation}\label{kcases}
\Psi(x)=\left\{\begin{array}{lr}
    \sin{x},~~~\OK^0<0\\
     x,~~~~~~~~\OK^0=0\\
    \sinh{x},~~\OK^0>0\,.
       \end{array}\right.
\end{equation}

Here the difference with respect to the constant CC case is
encoded in $H(z,\OM^0,\OL^0;\nu)$. For $\nu=0$,
$H(z,\OM^0,\OL^0;\nu=0)$ becomes the standard FLRW function
(\ref{HzSS}), and the luminosity distance (\ref{dist}) also
reduces to the standard form.

We use the magnitude data from the SCP (Supernova
Cosmology Project) \cite{p99}. The set includes 16 low-redshift
supernovae from the Cal\'an/Tololo survey and 38 high-redshift supernovae
(Fig.\ref{fig:mz}) used in the main fit of \cite{p99}.
Observational data are corrected such that they can be used as
the intrinsic magnitude of the object. The
effective value of the magnitude is obtained from that at the
peak of the light curve according to the expression:
\begin{equation} \label{effmag}
m_B^{eff} = m_X^{peak} + \alpha (s-1) - K_{BX} - A_X \equiv
m^{eff}+\alpha (s-1)\,.
\end{equation}
Here $\alpha$ is the parameter describing the correlation between
maximum brightness and rate of decline of  SNe Ia; $s$ is the
stretch factor mentioned above; X is the observed band; $K_{BX}$
is the correction to the change from the emitted B-band to the
received X-band, and $A_X$ is the galactic extinction.

%%%%%%%%%%%%%%%%%%%%%%%%%%%%%%%%%%%%%%%%%%%%%%%%%%%%%%%%%%%%%%%%%%%%%%%%
\figeight

%%%%%%%%%%%%%%%%%%%%%%%%%%%%%%%%%%%%%%%%%%%%%%%%%%%%%%%%%%%%%%%%%%%%%%%%

Figure \ref{fig:mz} represents, in the magnitude-redshift space,
the data obtained by the SCP Collaboration together with the
predicted magnitude-redshift relation {for our model based on the
scale (\ref{muH}) and two RG models based on the scale
(\ref{muCD})\,\cite{cosm,Babic}}. At low redshift all the models
are equivalent as it occurs with all the alternative models to the
standard one. At high redshift they display small variations with
respect to the ``constant CC'' cosmological model. As it is seen
graphically, current data are not able to distinguish between
models. At present, this kind of models can only be favored
theoretically (see Section \ref{sect:snap} for the future
prospects).

In order to determine the cosmological parameters we use a
$\chi^2$-statistic test, where $\chi^2$ is defined by the
difference between the theoretical apparent magnitude and the
observed one:
\begin{equation}\label{chi2}
\chi^2(\Omega_M,\Omega_\Lambda,\nu,\alpha,{\cal{M}})=\sum_i
\frac{\left\{{\cal{M}}
+5\log_{10}\left(H_0\,d_L(z_i,H_0,\OM,\OL,\nu)\right)-m_i^{eff}-
\alpha(s_i-1)\right\}^2}{\sigma_i^2}
\end{equation}
%and where the term $\alpha(1-s_i)$ calibrates the magnitude of the supernova
% to a fiducial light curve \cite{strech}.
As we are only interested in the cosmological parameters
associated to the acceleration of the universe, we marginalize
over $\alpha$ and ${\cal{M}}$ \cite{p99, astier}. That means that
we minimize $\chi^2$ by integrating over all possible values of
$\alpha$ and ${\cal{M}}$.

\subsection{Results from current data}
In this section, confidence levels surrounding best fits (minimum
$\chi^2$) are given by contour lines of constant $\chi^2$, which
represent
$1\sigma$, $2\sigma$ and $3\sigma$ confidence levels respectively.

A distinguishing feature of all the models is the evolution of
the cosmological constant, thus we can make a first test in order to
see whether such an evolution is consistent with current data. We
adopt a generic form for the Hubble parameter which describes any
CC running:
\begin{equation} \label{hubblevar}
 H^2(z)=H_0^2\,\left[\OM^0\,(1+z)^3 + \OL^0
+ \frac{1}{\rc^0}\left.\frac{d\CC}{dz}\right|_{z=0}\,z +
 \OK^0\,(1+z)^2\right]\,.
\end{equation}
This equation parameterizes generically the deviations from the
standard law (\ref{HzSS}) through a series expansion of
$\Lambda(z)$ up to first order in $z$, or equivalently up to the
term defining the $\beta$-function for each model - remember that
$d\CC/dz=\bCC\,d\ln H/dz$. We may therefore distinguish models by
plugging in the corresponding $\beta$-function in this formula.

In Fig.\,\ref{fig:confpla}a the confidence region in the
$(\Omega_\Lambda^0,({1}/{\rho_c^0}){d\Lambda}/{dz}\left|_{z=0}\right.)$
space for a flat universe is shown. It does by no means discard
the running of the cosmological constant. In some sense this is
similar to what happens with quintessence models where data are
compatible with a slow evolution of a scalar field, but here we
are testing the evolution of a density component without any
variation in the equation of state. When we restrict ourselves to
the constant cosmological term, $d\Lambda/{dz}=0$, we recover
results similar to the standard ones, namely
$\Omega_M^0\sim0.3\pm 0.1, \Omega_\Lambda^0\sim0.7\pm 0.1$ for a
flat geometry (Cf. Fig.\,\ref{fig:confpla}a).  In this case the
minimum $\chi^2$ value lies at
 $\Omega_M^0=0.29, \Omega_\Lambda^0=0.71$. But in general we
see from Fig.\,\,\ref{fig:confpla}a that this commonly accepted
value may undergo a wide variation when the possible running of
the cosmological constant is taken into account.

%\fignine
\figten

After this generic observation, we turn to our model and apply a
$\chi^2$ test to obtain confidence regions in the
$(\Omega_\Lambda^0, \nu)$-plane for a flat universe
(Fig.\,\ref{fig:confpla}b). We see there that $\Omega_{\Lambda}^0$
does not vary significantly with different values of $\nu$. This
is because the sample of supernovae used has redshift up to
$0.83$ (Cf. Fig.\,\ref{fig:mz})  and the average is $z\sim 0.5$.
However, as it was already observed in Fig.\,\ref{fig:evmupla1},
the effect becomes more important as we go to higher $z$, and we
will thus be able to draw more trustworthy conclusions when we
analyze SNAP data, which will reach up to $z=1.7$.

%%%%%%%%%%%%%%%%%%%%%%%%%%%%%%%%%%%%%%%%%%%%%%%%%%%%%%%%%%%%%%%%%%%%%%%
%\clearpage
\section{SNAP and the running cosmological constant}
\label{sect:snap}

After the discovery of the accelerated expansion of the
universe \cite{p99,Riess98} a new satellite observatory
(SuperNova Acceleration Probe)  was proposed  to determine
the nature of the dark energy cause of the acceleration. The SNAP
collaboration aims to obtain spectra and photometry for
 2,000 supernovae already in the first year of mission \cite{SNAP}. The
distribution of supernovae will have a maximum in the interval
$0.2<z<1.2$ where according to the present observed rates 1800
supernovae should be found. A smaller number of data is expected
to be obtained up to a redshift of 1.7 (see more details in
\cite{SNAP}).

We suppose that supernovae are equidistant within every redshift
interval, and obtain the magnitude for these supernovae within
the cosmological model presented in this work. In each case
observational gaussian errors are added to these values taking
into account the systematic uncertainties, which are estimated to
be zero at z=0 and 0.02 at z=1.5 following, for instance
\cite{weller}: $ \sigma_{sys}= ~z~ ({0.02}/{1.5}) $ and the
intrinsic dispersion of supernovae after the corresponding
calibrations $\sigma_{intr}=~0.15$.

After fitting the models to their simulated data we  have an
idea of the precision with which we will determine the
parameters and the possibility of distinguishing among models.
 Moreover, when the first results of SNAP
will become available, the value of $\Omega_M^0$ should already
be known to a high precision. We can therefore adapt the present
study to this foreseeable situation and use a Bayesian analysis
in order to determine the probability of our parameters given
some  prior information. Details of the statistical methods used
are shown in Appendix 2.

\subsection{High--z samples and the determination of $\nu$}
With the SNAP sample we will have, at high redshift, groups of 20
supernovae per bin of width 0.01 in redshift. This means that the
uncertainty in the magnitude at each bin will be of the order of
only 0.03 magnitudes for the present intrinsic dispersion of 0.15
magnitudes.

We can take one of this bins at relatively low redshift
($z_1=0.5$) and another one at higher redshift
($z_2=1.0\,$ or $\,1.5$)  to
look at
 the intersection between the allowed regions in the parameter space
 (Fig.\,\ref{fig:rombe}).
This intersection represents how the intrinsic dispersion in
magnitude translates into the parameters according to ($
\Delta\Omega_\Lambda \times \Delta\Omega_M ) \propto
(\sigma_m^{z_1} \times \sigma_m^{z_2}) $ (see for instance
\cite{p95}).

\figeleven

Figure \ref{fig:rombe} shows the results when the running of the
cosmological constant is considered. In the left panel the bands
of constant magnitude are plotted in the usual space of
cosmological parameters. The cosmological model used is one with
$\Omega_M^0=0.3, \Omega_\Lambda^0=0.7, \nu=2\nu_0$ and the bands
give a similar intersection as in a standard model with
$\Omega_M^0=0.3, \Omega_\Lambda^0=0.7$ (Fig.\,2 in \cite{p95}), so
the present-day parameters will be determined under the same
conditions in both kinds of models. But we can also see how the
cosmological index is determined by the SNAP measurements. In
order to obtain a reduced confidence region in the
($\Omega_\Lambda^0,\nu$) plane, we need data more separated in
redshift than in the previous case. This is because the
interesting models with standard (non-variable) cosmological
constant have their maximum difference at $z\sim0.8$ (this value
is that obtained for the difference between a $\Omega_M^0=0.3,
\Omega_\Lambda^0=0.7$ universe and a $\Omega_M^0=0.2,
\Omega_\Lambda^0=0$ universe). Models with a running cosmological
constant differ from the ones without such running CC
 at higher redshift ($z>1$) and the difference grows with $z$.
Thus, as we see in Fig.\,\ref{fig:riess}, we need a sizeable group
of supernovae at the highest redshift reachable
 by SNAP while a strong concentration of
data around $z=1$ is needed to distinguish
models with and without cosmological constant.

\figtwelve

\subsection{Numerical simulations for SNAP and other high--z SNe Ia samples}

 We now use the distribution of data predicted by the SNAP collaboration
%showed in Table \ref{tab:distrSNAP}
and some alternative distributions to obtain the confidence
regions for the representative parameter of the model, $\nu$.
However, as we have done with the current data, we first use the
distinctive characteristic of the variable CC models, namely the
presence of a term ${d\Lambda}/{dz}$ different from zero. The
order of magnitude of these derivatives in our model model can be
directly read from Fig.\, \ref{fig:evmupla2}d. At the present
epoch $({1}/{\rho_c^0}){d\Lambda}/{dz} = 3\nu\OM^0 \sim 0.05 $ for
$\nu=2\nu_0$. As seen in the figures, Eq. (\ref{hubblevar}) is not
a sufficiently good approximation in this case since the Taylor
expansion is only significant at first order. Therefore, in our
model we would have to use the second derivative as well in order
to study the change of slope at high redshift. The other
parameters used in the simulation are a flat universe with
$\Omega_M^0=0.3, \Omega_\Lambda^0=0.7$.

For such a universe we obtain, after the SNAP project, the
results in Fig.\,\ref{fig:derlambSNAP}. The left panel shows the
confidence regions in the
($\Omega_\Lambda^0,({1}/{\rho_c^0}){d\Lambda}/{dz}\left|_{z=0}\right.$)
plane when a flat universe is considered, and the right panel adds
a gaussian prior on $\Omega_M^0$ with $\sigma_{\Omega_M}=0.03$.
These restrictions in the analysis are justified by recent
results of the WMAP experiment \cite{WMAP}.
 Thus, with all these considerations,
 $({1}/{\rho_c^0}){d\Lambda}/{dz}$ at the present epoch
would be obtained to within $\pm0.2$  which could distinguish
models like the one presented here from the standard cosmological
constant model.

\figthirteen

The results for the parameter $\nu$ are shown in
Fig.\,\ref{fig:confSNAP}. Fits are made under the same
assumptions as in the previous case and so the results represent
the expected accuracy in the parameters using all the information
to come. Thus we determine $\nu$ to $\pm0.06$ for $\nu=0.1$.
 We must take into account, however,
that the SNAP distribution of data is optimized in order to
distinguish models that have their maximum difference at a
redshift near 1. We can, therefore, try other distributions
designed from the idea that the difference between models grows
with redshift, as it indeed happens in our case. The first
distribution is very similar to the SNAP one but with most of the
data homogeneously distributed between $z=0.2$ and $z=1.7$. This
second distribution extends data up to redshift $z=2$. This is in
part justified by data which are being obtained within the GOODS
and HST Treasury Program \cite{goods}. \figfourteen The results
from the various distributions are shown in
Fig.\,\ref{fig:nu_distr} and quantified in
Table\,\ref{tab:resultats}. The latter also compares the
determination of the parameters representing an RG evolution of
the cosmological constant (i.e. $\nu\neq 0$), and the evolution of
the equation of state (see Eq.\,(\ref{wexpansion})) for an
alternative dark energy model.
 We see that the evolution of the cosmological
constant can be determined with a relative error of $60\%$ for a
value of $\nu=0.1$. Distributions of data other than SNAP
(Distr.1 and Distr.2 in Table \ref{tab:resultats}) allow to
reduce this error to  $20\%$ with $\nu=0.1$, but determinations
remain poor for small values of $\nu$, which on the other hand
would be the natural ones. We are then in a situation similar to
the determination of the evolution of the equation of state,
Eq.\,(\ref{wexpansion}). Although $w_0$ can be obtained with a
precision of $2\%$ for $w_0 \sim -1$, its evolution $w_1$ is only
determined with great uncertainty. For instance, the evolution of
$w$ for SUGRA described by \cite{linder} is expected to be
determined with a relative error in $w_a \sim  2 w_1$ of 43 $\%$.
In our case, as for other dark energy candidates, very high--z
samples as the ones being collected by GOODS  and HST Treasury
program will significantly contribute to the precise
determination of $\nu$.

\figfifteen

 \vspace{0.5cm}

\TABULAR[h]{l|c|c|c|c} { \hline \hline
              &      &           & \\
 Distribution~~        & Data      & Prior $\sigma_{\Omega_M}$ &
        $\theta$& $\sigma_\theta$  \\
              &      &           & \\
\hline
               &                   &      &           &             \\
{\bf SNAP}     & 50 SNe $0<z<0.2$  &      &           &             \\
(1 year)      &1800 SNe $0.2<z<1.2$&      &           &             \\
               &50 SNe $1.2<z<1.4$ &      &           &             \\
               &15 SNe $1.4<z<1.7$ & None & $\nu=0.1$ &~~$\pm0.10$  \\
               &                   &      &           &             \\
{\bf SNAP}     & as above          & 0.03 & $\nu=0.1$ &~~$\pm0.06$  \\
               &                   &      &           &             \\
{\bf SNAP}     &  3 years          & None & $\nu=0.1$ &~~$\pm0.06$  \\
               &                   &      &           &             \\
{\bf SNAP}     &  3 years          & 0.03 & $\nu=0.1$ &~~$\pm0.04$  \\
               &                   &      &           &             \\
{\bf Distr.1}  & 50 SNe $0<z<0.2$  &      &           &             \\
             &2000 SNe $0.2<z<1.7$ & 0.03 & $\nu=0.1$ &~~$\pm0.05$  \\
               &                   &      &           &             \\
{\bf Distr.2}  & 250 SNe $0<z<1$   &      &           &             \\
               &1750 SNe $1<z<2$   & 0.03 & $\nu=0.1$ &~~$\pm0.02$  \\
               &                   &      &           &             \\
%\hline
%               &                   &      &           &             \\
%{\bf SNAP}   & & None & $\beta_\Lambda (10^{-9}eV^4)=10$ & ~~$\pm4$ \\
%               &                   &      &           &             \\
%{\bf SNAP}   & & 0.03 & $\beta_\Lambda (10^{-9}eV^4)=10$ & ~~$\pm2$ \\
%               &                   &      &           &             \\
\hline
               &                   &      &           &             \\
{\bf SNAP}     &  3 years           & 0.03 & $w_0=-1$    & ~~$\pm 0.02 $ \\
               &                   &      &           &             \\
{\bf SNAP}     &                   &       & $w_a=0.58 $ & ~~$\pm 0.25$ \\
               &                   &      &           &             \\
\hline \hline
 }
{ Determination of the parameters with SNAP data and with other
two distributions. In all cases the cosmology used for the
analysis is a flat universe, and when a prior on $\Omega_M$ is
added we use a central value of $\Omega_M=0.3$. For comparison we
show the parameters representing variations in the equation of
state (\ref{wexpansion}) (for SUGRA in \cite{linder}).
\label{tab:resultats}}

%\clearpage
\section{Conclusions}

\label{sect:conclusions}

To summarize, we have constructed a semiclassical FLRW model with
variable CC at the present cosmic scale. The variation of the
vacuum energy is provided without introducing special scalar
(quintessence-like) fields and is completely caused by quantum
effects of vacuum. The evolution of the CC is due to the
Renormalization Group running triggered by the smooth decoupling
of the massive fields at low energies, while the RG scale $\mu$
being associated to the Hubble parameter, $H$, at the
corresponding epoch. Although the $\bCC$-function itself is
proportional to the fourth power of the masses, the decoupling
does still introduce an inverse power suppression by the heavy
masses, and thus one is left with a residual quadratic law
$\bCC\sim H^2 M^2$. The effective scale $M$ summarizes the
presence of the heaviest degrees of freedom available. This
peculiar form of decoupling stems directly from: \,i) the
Appelquist-Carazzone (AC)-decoupling theorem, \,ii) general
covariance, and also from\, iii) the \textit{non}-fine-tuning
hypothesis between the $n=1$ (``soft decoupling'') terms of
$\bCC$ (Cf. Eq. (\ref{newRG1})) whereby the overall coefficient
of the quadratic contribution $H^2 M^2$ does not vanish. This
particular form of decoupling is a specific characteristic of the
CC because it is of dimension four. There is no other parameter
either in the SM or in the GUT models with such property.

In constructing the cosmological model we have explored the
possibility that the heaviest d.o.f. may be associated to
particles having the masses just below the Planck scale. This
assumption is essential to implement the soft decoupling
hypothesis within the $\mu=H$ setting. Indeed,  the present value
of $\,H_0^2 M^2\,$ is just of the order of the CC, and therefore
it insures a smooth running of the cosmological term around the
present time. In this setting the $\bCC$-function has only one
arbitrary parameter $\,\nu\,$ (\ref{nu}), and as a result the
model has an essential predictive power. In general we expect
$\,|\nu|\ll 1\,$ from phenomenological considerations, mainly
based on the most conservative hypotheses on the nucleosynthesis
framework. Furthermore, for $\,|\nu|\ll 1$, we insure the absence
of the trans-Planckian energies. It is not completely clear how
much the cosmological index $\,\nu\,$ can approach the value $1$
from below. Despite the values $\nu\lesssim 1$ are not completely
ruled out, it is not clear that the conditions for the
nucleosynthesis could be safe. However, if accepted, it opens up
the possibility that the CC could be in transit from a $\,\CC>0\,$
regime into a future $\,\CC\leq 0\,$ one, which would be a
welcome feature for string/M-theory.

Our RG model offers the possibility to explore the existence of
the sub-Planck physics in direct cosmological experiments, such
as SNAP (and the very high--z SNe Ia data to be obtained with
HST). For example, for the flat FLRW case and a moderate and
positive value $\nu \sim 10^{-2}$, we predict an increase of
$10-20\%$ in the value of $\Omega_{\Lambda}$ at redshifts
$z\gtrsim 1-1.5$ perfectly reachable by SNAP. For similar, but
negative, values of $\nu$ we predict that the CC should become
negative beyond redshifts $z\gtrsim 2$. For $\nu\sim 0.1$
corrections to some FLRW cosmological parameters become as large
as $50\%$ or more. In general, this model has a wide spectrum of
implications that could be tested by SNAP, even for fairly
moderate values of $\nu$ compatible with the most conservative
bounds from nucleosynthesis. The simulations of the SNAP data in
Section 7 show that the single parameter $\nu$ of our model could
be pinned down, for $\nu\lesssim 0.1$, with a precision of around
$20-60\%$. Although this accuracy is not very high, the sign of
the parameter could be determined and the effects would be
manifest. Undoubtedly, this would be a good starting point to
identify the presence of quantum corrections to the FLRW
classical cosmology.

The semiclassical FLRW model that we have proposed here explains
the variation of the CC due to the ``relic'' quantum effects
associated to the decoupling of the heaviest degrees of freedom
below the Planck scale, and suggest that a time dependence of the
CC may be achieved without resorting to scalar fields mimicking
the cosmological term or to modifications of the structure of the
SM of the strong and electroweak interactions and/or of the
gravitational interactions. At the same time, our model provides a
phenomenological parametrization for possible correlated
deviations of the cosmological equations for matter density and
dark energy.
%%%%%%%%%%%%%%%%%%%%%%%%%%%%%%%%%%%%%%%%%%%%%%%%%%%%%%%%%%%%%%%%%%
% In other words, even in the absence of a fundamental
% explanation for the cosmological index $\nu$ in Eq. (\ref{nu}),
% the extensive numerical analysis that we have performed is useful
% to test the potential deviations of the cosmological laws from
% the standard FLRW frame. Within this purely phenomenological
% framework we have seen that the allowed departures are not
% just mere one per mil corrections but much larger, reaching the
% level of $10\%$ or higher.

From a more fundamental point of view, we have shown that the
cosmological constant and the matter density may evolve in a
correlated way due to quantum effects without resorting to
exceedingly exotic frameworks.
%%%%%%%%%%%%%%%%%%%%%%%%%%%%%%%%%%%%%%%%%%%%%%%%%%%%%%%%%%%%%%
% The cosmological model presented here is, theoretically,   %
% of a fairly wide scope.                                    %
% The $\,H^2$-type quantum corrections to the vacuum energy  %
% are not specific to a given QFT or string-like context,    %
% indeed they could emerge as an effective framework common  %
% to the physics of a large class of theories defined near   %
% the Planck scale.                                          %
%%%%%%%%%%%%%%%%%%%%%%%%%%%%%%%%%%%%%%%%%%%%%%%%%%%%%%%%%%%%%%
Using the RG as a basic QFT tool and extrapolating the standard AC
law of decoupling to the CC case, we have straightforwardly
predicted a cubic redshift evolution law for the CC and the
matter density. The larger the redshift that we can eventually
explore the larger the effects that we predict. Most important,
we have shown that this cubic law can be thoroughly tested by the
next generation of cosmological measurements, which will be able
to reach depths up to $z\sim 2$.

%%%%%%%%%%%%%%%%%%%%%%%%%%%%%%%%%%%%%%%%%%%%%%%%%%%%%%%%%%%%%%%%%%
\acknowledgments
 C.E.B and P.R.L are partially supported by a European Research and
 Training Network Grant on Type Ia Supernovae (HPRN--CT--20002-00303),
 and by research grants in cosmology by the Spanish DGYCIT
(ESP20014642--E) and Generalitat de Catalunya
 (UNI/2120/2002). P.R.L. thanks the support and hospitality
 of the MPA
 in Garching. I. Sh. and J.S. are thankful to E.V. Gorbar,
 B. Guberina, M. Reuter, H. Stefancic and A. Starobinsky for fruitful
 discussions. The work of I.Sh. has been supported by
the research grant from FAPEMIG (MG, Brazil) and by the fellowship
 from CNPq (Brazil).
 I.Sh. also thanks the Erwin Schrodinger
 Institute in Vienna and the Department of Theoretical Physics
 at the U. Zaragoza and the Dep. E.C.M at the U. Barcelona
 for the kind hospitality and support.
 The work of J.S. has been supported in part by MECYT and FEDER
 under project FPA2001-3598, and also by the Dep. de Recerca de la
 Generalitat de Catalunya under contract 2002BEAI400036. J.S. is
 also thankful to the Dep. de Fisica UFJF Brazil and to the MPI Munich
 for the hospitality and
 financial support.

%%%%%%%%%%%%%%%%%%%%%%%%%%%%%%%%%%%%%%%%%%%%%%%%%%%%%%%
\section{Appendix 1. On the non-local terms in the
quantum corrections.}
\label{sect:A}

In this appendix we shall go beyond the phenomenological
presentation used throughout the main text and briefly discuss
some theoretical aspects of our QFT framework. Despite the
cosmological model developed in the main text of the paper looks
appealing due to its relative simplicity and relation to the QFT
corrections, it is not immediately clear which level of
credibility can be granted to the hypothesis of the ``soft''
low-energy decoupling which we used here. As it was already
remarked in the main text, the physical interpretation of the
renormalization group in curved space-time requires the formalism
beyond the limits of the well-known standard techniques. These
techniques are essentially based on the minimal subtraction
$\,\overline{MS}\,$ scheme of renormalization \cite{Collins,
LBrown, tmf,book}, which does not admit to observe the decoupling.

The physical interpretation of the renormalization group in the
higher derivative sector can be achieved through the calculation
of the polarization operator of gravitons arising from the matter
loops in linearized gravity \cite{apco}. In this case one can
perform the calculations in the physical mass-dependent
renormalization scheme (specifically, in the momentum subtraction
scheme, in which $\mu$ is traded for an Euclidean momentum $p$),
obtain explicit expressions for the physical beta-functions and
observe the decoupling of massive particles at low energies. In
the mass-dependent scheme one has direct control of the
functional dependence of the $\,\beta$-functions on the masses of
all the fields, i.e. one knows explicitly the functions
$\,\beta_k(m_i;p^2/M^2_j)\,$ for the higher derivative terms $a_k$
in (\ref{Svac}). Quite reassuring is the fact that these momentum
subtraction $\,\,\beta$-functions boil down, in the UV limit, to
the corresponding $\,\,\beta$-functions in the MS scheme
\cite{apco}.

Unfortunately, as it was realized in \cite{apco}, this program
fails when we attempt to derive the physical $\,\beta$-function
for the CC (and also for the inverse Newton constant $\,1/G$). In
fact, within the momentum subtraction scheme we fail to see the
$\,\beta$-function for CC and for the inverse Newton constant
$\,1/G\,$ at all, so there is no correspondence with the
$\,\overline{MS}\,$ scheme in these sectors. The origin of the
problem is that the standard renormalization group running always
corresponds to the non-local insertions in the effective action.
For example, in case of QED the one-loop terms look, in the UV,
like
 \beq \Gamma_{QED} \,=\,\int d^4x \,\sqrt{-g} \,\,\Big\{\,\,
-\frac{1}{4}F_{\mu\nu}F^{\mu\nu} + \frac{e^2}{3(4\pi
)^2}F_{\mu\nu}\, \log
\Big(\frac{\square}{m_e^2}\Big)\,F^{\mu\nu}\,\Big\}\,. \label{Z3}
\eeq
 In the $\overline{\rm MS}$-scheme of
renormalization the total $\mu$-independence of the effective
action (see, e.g. \cite{book}) is provided by the running of the
coupling $\,e=e(\mu)$. This running is a physical effect which
has been observed experimentally \cite{ParticleData}. As noted
before, if we wish to describe low energy effects $p^2\ll m_e^2$,
we need a more physical subtraction scheme where the role of
$\mu$ is played by the momentum $p$. Then we expect decoupling
terms of the form $p^2/m_e^2$, which manifest as additional
$\square/m_e^2$ terms (without logarithms) in the effective
action (\ref{Z3}). Similarly, in the higher derivative
gravitational sector we have the terms of the form \beq
\Gamma_{HD}\,=\,\int d^4x \,\sqrt{-g} \,\,\big\{\,
C_{\mu\nu\alpha\beta} \,k_1(\square,m^2)\,C^{\mu\nu\alpha\beta}
\,+\,R \,k_2(\square,m^2)\,R\,\big\}\,, \label{HDG} \eeq where
$\,C_{\mu\nu\alpha\beta}\,$ is the Weyl tensor, $\,R\,$ is a
scalar curvature, $m$ is any particle mass and $\,k_1\,,\,k_2\,$
are definite non-local functions \cite{apco} which in the high
energy region include $\,\ln(\square/m^2)$. As in the QED case,
at low energy these functions contain the decoupling terms
$\square/m^2$ without logs. When the derivative operators act on
the conformal factor $a(t)$, they produce terms proportional to
the powers of the Hubble parameter $H$ and its derivatives. Since
the derivative ${\dot H}$ of the Hubble parameter has the same
order of magnitude as $H^2$, for the sake of simplicity we can
consider only the powers of $H$. It is clear that the finite order
non-localities can be described, at low energies, by the local
expansions with the tensor structure similar to the one of
(\ref{HDG}). For instance, making the expansion in the powers of
$\,\square/m^2$, we meet the quadratic law of the low-energy
decoupling.

Let us now consider the cosmological and Einstein-Hilbert terms.
In the $\overline{\rm MS}$-scheme the $\mu$-dependence of the
form of the effective action is compensated by the running of the
parameter $\,\Lambda$, similar to the QED case and the running
$e(\mu)$. As a result the overall effective action with a running
CC is scale-independent, as it has to be. We are able, in
principle, to observe the CC only at very small energies, and
naturally expect to have much weaker running because of the
low-energy decoupling. However, the quantitative investigation of
the phenomena of decoupling for the CC met very serious
difficulties, starting from the physical interpretation of the
renormalization group in this case. As noticed in \cite{apco},
the non-local insertions can not be done in the original
gravitational terms, because acting on $\CC$ by $\,\square\,$ we
get zero and $\,\square\,$ acting on $\,R\,$ gives a total
derivative. The insertion can be done only into the non-local
terms with similar global scaling, e.g. \beq \int d^4x
\sqrt{-g}\, R_{\mu\nu}\Big(\frac{m^2}{\square}\Big)^2\, R^{\mu\nu}
\qquad \mbox{and} \qquad \int d^4x \sqrt{-g}\,R_{\mu\nu}
\Big(\frac{m^2}{\square}\Big)R_{\mu\nu} \label{h} \eeq for the CC
and inverse Newton constant respectively. For the effective
action of the quantized massive fields these terms do not take
place and hence the framework of linearized gravity is not
appropriate for the physical interpretation of the
renormalization group for the CC and $\,1/G\,$ in curved space.
Does it mean that there is no other framework where this
interpretation should appear in a natural way? From our point of
view it is not really so. Let us remind that: \vskip 1mm

1. $\,$ The renormalization group for the CC and $\,1/G\,$ can be
formulated in a very robust way in curved space if we use the
$\,\overline{MS}$ renormalization scheme. Therefore, we can
expect the coincidence with the physical RG in an appropriate
framework\footnote{Let us notice that the renormalization group
for the CC and $\,1/G\,$ is nontrivial within the
non-perturbative approach -- see, e.g. \cite{Reuter03a}. This
supports our point of view on the scheme-dependence of the
negative results of \cite{apco}.}. \vskip 1mm

2. $\,$ The linearized gravity approach is essentially based on
the expansion near the flat background. But when quantizing
massive fields we have to introduce the vacuum CC from the very
beginning in order to provide renormalizability. In this case the
flat Minkowski space-time is not a solution of the classical
equations of motion and the whole approach based on the
linearized gravity is formally illegal. Perhaps this is the
source of the problem. \vskip 1mm

In order to see the renormalization group and decoupling for the
CC and $\,1/G\,$ one has to perform calculations of the
polarization operator or vertices on the non-flat dynamical
background. In this case there will be no relation between the
covariant expansion in curvatures and the gravitational
perturbations around the given metric. From the point of view of
the linearized gravity this means that we can not achieve the
physical interpretation of the renormalization group via the
gravitational perturbations around the flat metric background,
because all the effect is essentially non-perturbative. In some
sense the situation is similar to the low-energy QCD, where one
has to rely on the non-perturbative methods or reformulate the
perturbative calculations using, e.g. the renormalon technique.
It might happen that the resummation of the perturbative series
coming from both quantum loop expansion and the gravitational
perturbations may lead to the visible renormalization group
effect, but at present it is unclear how this could be checked.

Let us consider again the linearized gravity approach.
The non-localities of the effective action
of the massive fields are related to the insertions
of the ``massive'' Green functions
$\,{1}/(\square+m^2)\,$. Then, making the expansion
in the powers of $\,\square/m^2$, we meet, typically,
much stronger decoupling than the standard quadratic
one. However, since we are dealing with an infinite
multiple series, the correspondence with the
$\,\overline{MS}$ scheme means that some resummation
of the series in curvature produces the massless
$\,\,{1}/{\square}\,$ insertion which may eventually
lead to the soft decoupling and to the UV correspondence
with the $\,\overline{MS}$ scheme.

At present there is no method of calculations on the non-flat
background compatible with the physical renormalization scheme.
In this situation our phenomenological approach looks rather
justified. The covariance of the effective action forbids the
first order in $\,H\,$ corrections. Then, recalling that the UV
contribution to $\,\beta_\Lambda\,$ from a particle of mass
$\,m\,$ is $\beta_\Lambda\sim m^4$ -- Cf. Section \ref{sect:RCC}--
the low-energy contributions to $\,\beta_\Lambda\,$ must be
supressed by, at least, the factor
\begin{equation}
\square/m^2\,\,\sim\,\,H^2/m^2\,, \label{endequation}
\end{equation}
and hence the overall low-energy $\,\beta_\Lambda\,$ acquires the
form $H^2\,m^2$, Eq.\,(\ref{RG1}). The same form of decoupling
can be expected for the induced counterpart $\CC_{ind}$ of the
CC. The renormalization group equations for the vacuum and
induced CC's are independent, and the relation emerges only at
the moment when we choose the initial point of the RG trajectory
for the vacuum counterpart \cite{JHEPCC1,cosm}. In fact, the
Feynman diagrams corresponding to the $\CC_{ind}$ are, at the
cosmic scale, very similar to the vacuum ones, because the Higgs
field $\phi$ is in the vacuum state, $<\phi>\neq 0$, and the
non-zero contribution of these diagrams is exclusively due to the
momenta coming from the graviton external lines. Finally, there
is no real need to distinguish induced and vacuum CC's in the
present context. In both cases the absence of the $n=0$ order
(non-supressed) contributions in (\ref{newRG1}) is required by
the apparent correctness of the Einstein equations and the
smallness of the observable CC. There is no guarantee that the IR
supression does not take a much stronger form than the one (soft
quadratic decoupling) assumed here. In the last case the future
experiments and observations will demonstrate a perfectly
constant\,CC\footnote{In case there is an additional source of
the dark energy (e.g. quintessence), the variable dark energy
will manifest itself.}. The same should happen if the
high-energy spectrum does not include the particles with the
masses comparable to $\,M_P$. In both these cases the effect of
running would be negligible. Then, since our
renormalization-group based model gives very definite prediction
for many observables (see Sections 4,5), it should be, sooner or
later, distinguished from the standard model with the
$\,z$-independent CC or maybe even from the quintessence models.

One could wonder whether the non-local effects behind the
renormalization group are compatible with the standard energy
conservation equation which we used in the text. In fact, the
covariant form of the conservation law $\,<\nabla_\mu T^\mu_\nu >
=0\,$ just reflects the covariance of the effective action and
therefore does not depend on the non-localities which are always
present in the quantum corrections. Furthermore, in a situation
where the energy scale associated to the metric derivatives is
very small, the effect of non-local terms can be, according to
our model, presented in a compact form due to the renormalization
group. In the leading order of magnitude one can use the
renormalization-group induced terms to represent the leading
non-local part of the effective action. In this case these terms
should be many orders of magnitude larger than the ones
corresponding to the next order approximation, and the
conservation law should be valid especially for the
renormalization group induced terms.

Let us finally remark on the running in other sectors. For the
inverse Newton constant $1/G\,$ it is irrelevant because $1/G\sim
M_P^2$ is very large and the effect of the running is relatively
small {as it was demonstrated in \cite{JHEPCC1}}. The relevance of
the higher derivative terms at low energies is supposed to be
negligible, as it was recently shown in \cite{Pelinson} for the
perturbations of the conformal factor. Indeed, for the tensor
perturbations the situation may be much more complicated (see,
e.g. \cite{Mottola} and references therein). In contrast, the
soft decoupling for the CC does matter because the CC is very
small and any running, even the very small one that we get, is
sufficient to produce some measurable effect at large redshifts.
For precisely this reason we have to use the Friedmann equation,
in combination with the full conservation law including the
variable cosmological term, and no other terms.

\section{Appendix 2. Bayesian analysis.}
\label{sect:A2}

In the Bayesian approach we can determine the probability density
function for the cosmological parameters
($\Omega_M,\Omega_\Lambda$) and model parameters ($\vec{\theta}$
in general), conditional on the data ($\vec{D}$). We know from
 Bayes's theorem that this quantity,
$p(\Omega_M,\Omega_\Lambda,\vec{\theta}|\vec{D})$, is obtained
from the probability of the data conditional on the model and
from the a priori information on the parameters and on the data:

\begin{equation}
p(\Omega_M,\Omega_\Lambda,\vec{\theta}|\vec{D})=
\frac{p(\vec{D}|\Omega_M,\Omega_\Lambda,\vec{\theta})
~p(\Omega_M,\Omega_\Lambda,\vec{\theta})}{p(\vec{D})}.
\end{equation}

Data coming from supenovae are effective magnitudes in the peak
of the light curve, $m_i^{eff}$ given in Eq. (\ref{effmag}), and
in this analysis $\vec{\theta}=({\cal{M}},\alpha,\nu)$.

We assume that $m_i^{eff}$ are independent and each one follows a
gaussian distribution. Then we obtain a probability density
function for the data conditional on the model that is a product
of gaussians. Hence the exponent is proportional to the sum that
we have defined as $\chi^2$ in Eq. (\ref{chi2}),

\begin{equation}
p(m_i^{eff}|\Omega_M,\Omega_\Lambda,\vec{\theta})=
\prod_i\frac{1}{\sqrt{2\pi\sigma_i^2}}
\exp\left(-\frac{1}{2}\frac{(m_i^{th}-m_i^{eff})^2}{\sigma_i^2}\right)
\propto \exp\left(-\frac{\chi^2}{2}\right)\,.
\end{equation}

As in the $\chi^2$ analysis we follow the results from
\cite{astier} and we reduce $\vec{\theta}$ only to the parameter
of the model ($\nu$ in our case). Since we concentrate on the
flat case we only introduce a gaussian prior on $\Omega_M$. Using
that $p(m_i^{eff})$ is constant too we obtain

\begin{equation}
p(\Omega_M)=\frac{1}{\sqrt{2\pi\sigma_{\Omega_M}^2}}\exp
\left(-\frac{1}{2}\frac{(\Omega_M-\Omega_M^0)^2}{\sigma_{\Omega_M}^2}\right),
\end{equation}

\begin{equation}
p(\Omega_M,\Omega_\Lambda,\nu |m_i^{eff})=\frac{1}{N}
 \exp\left(-\frac{\chi^2}{2}\right)\exp\left(-\frac{1}{2}
\frac{(\Omega_M-\Omega_M^0)^2}{\sigma_{\Omega_M}^2}\right),
\end{equation}
where N is the normalization given by

\begin{equation}
N=\int_0^\infty d\Omega_M \int_{-\infty}^\infty d\Omega_\Lambda
 \int_{-\infty}^\infty d\nu
 \exp\left(-\frac{\chi^2}{2}\right)\exp\left(-\frac{1}{2}
\frac{(\Omega_M-\Omega_M^0)^2}{\sigma_{\Omega_M}^2}\right).
\end{equation} \\

%\newpage
%%%%%%%%%%%%%%%%%%%%%%%%%%%%%%%%%%%%%%%%%%%%%%%%%%%%%%

%%%%%%%%%%%%%%%%%%%%%%%%%%%%%%%%%%%%%%%%%%%%%%%%%%%%%%%%%%%%%%%%
\end{document}